# Plasmonic Metamaterials

Kan Yao[2] and Yongmin Liu[1,2,*]

[1]Department of Mechanical and Industrial Engineering, Northeastern University,
Boston, MA 02115, USA

[2]Department of Electrical and Computer Engineering, Northeastern University,
Boston, MA 02115, USA

*Corresponding author: y.liu@neu.edu

## Abstract

Plasmonics and metamaterials have attracted considerable attention over the past decade, owing to the revolutionary impacts that they bring to both the fundamental physics and practical applications in multiple disciplines. Although the two fields initially advanced along their individual trajectories in parallel, they started to interfere with each other when metamaterials reached the optical regime. The dynamic interplay between plasmonics and metamaterials has generated a number of innovative concepts and approaches, which are impossible with either area alone. This review presents the fundamentals, recent advances and future perspectives in the emerging field of plasmonic metamaterials, aiming to open up new exciting opportunities for nanoscience and nanotechnology.

## 1.  Introduction

Since the early 2000s, metamaterials have emerged as a new frontier of science involving physics, material science, engineering, optics and nanoscience. The primary reason for the extensive interest in metamaterials lies in the fact that metamaterials have implemented a wide range of exceptional properties by engineering the internal physical structures of their building blocks. This is distinctly different from natural materials, whose properties are primarily determined by the chemical constituents and bonds.

Metamaterials comprise periodically or randomly distributed artificial structures with the size and spacing much smaller than the wavelength of interest.[1-4] From the deep subwavelength feature, the microscopic detail of each individual structure cannot be resolved by the electromagnetic waves. We can homogenize the assembly of these inhomogeneous elements, and assign effective material properties at the macroscopic level.[5] Importantly, the effective material properties are predominantly determined by the size, shape or structure of the building blocks of metamaterials, instead of the intrinsic properties of the constituent materials that are used to construct metamaterials. Using different metamaterial designs (Figure 1), researchers have been able to engineer the material properties with unprecedented degrees of freedom, and have demonstrated extremely low-frequency plasmons,[6] artificial magnetism,[7] negative refractive index,[8] extremely large refractive index[9] and strong chirality.[10] The broad spectrum of material properties offered by metamaterials also propels the rapid development of transformation optics, which enables us to manipulate the flow of electromagnetic waves in almost arbitrary manners.[11-15]

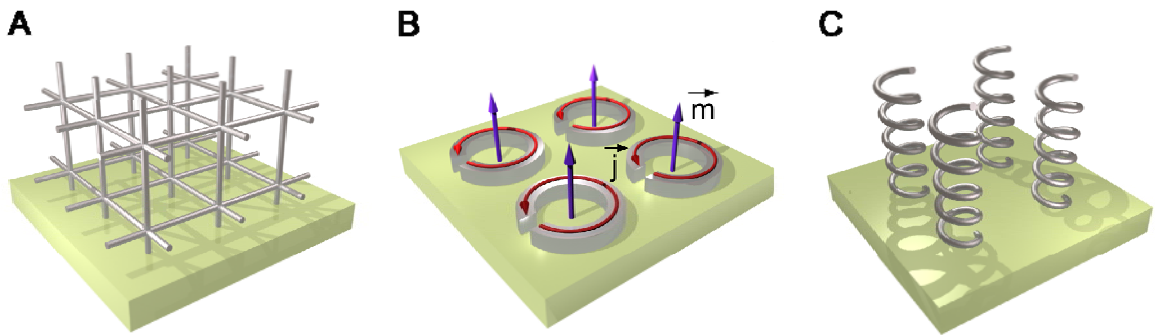

**Figure 1** Schematics of some representative metamaterials. (A) Metallic wires to achieve low-frequency plasmons by reducing the electron density and increasing the effective electron mass. (B) Metallic split ring resonators (SRRs) to produce artificial magnetism. Around the resonance frequency, there is a strong current ($\vec{j}$) circulating along the resonator, resulting in an effective magnetic moment ($\vec{m}$). (C) Chiral metamaterials made of metallic helices.

After the proof-of-principle experiments in microwave regime, the operation frequency of metamaterials was quickly pushed to higher frequency from terahertz (THz) to mid-/near-infrared and to visible.[16] Because most metamaterials are composed of metals, the plasmonic effect of metals plays an important role in optical metamaterials. For example, the magnetic resonance frequency of split ring resonators (SRRs) inversely scales with the structural dimension when the operation frequency is below ~100 THz, where the metal can still be treated as a perfect metal with infinite carrier density and zero carrier velocity.[4,17] However, this scaling law breaks down when approaching the optical regime and the model for real metals must be adopted. At optical frequencies, the kinetic energy of electrons can no longer be neglected comparing with the magnetic field energy driving the current in the loop, which contributes an additional term to the inductance of SRRs.[18,19] This electron inertia together with other plasmonic effects changes the simple scaling law in a complicated way, leading to an ultimate saturation of resonance frequencies of SRRs at about several hundred THz. Therefore, a more comprehensive exploration of plasmonic effects in metals will construct a solid foundation to develop optical metamaterials.

Plasmonics research focuses on the unique properties and applications of surface plasmon polaritons (SPPs), quasiparticles arising from the strong interaction between light and free electrons in metals.[20,21] At the interface between a semi-infinite metal and a semi-infinite dielectric, SPPs behave as a surface wave that propagates along the interface while exponentially decays into both the dielectric and metal (Figures 2A and 2B). The dispersion relation of SPPs at a dielectric-metal interface can be obtained by solving Maxwell's equations and applying proper boundary conditions. Denoting the dielectric constant of the metal and the dielectric material as $\varepsilon_m$ and $\varepsilon_d$, respectively, the dispersion relation of surface plasmons is written as

$$k_z = \frac{\omega}{c}\sqrt{\frac{\varepsilon_m \varepsilon_d}{\varepsilon_m + \varepsilon_d}} \qquad (1)$$

where $\omega$ is the angular frequency, $c$ is the speed of light in vacuum, and $k_z$ is the wave vector of SPPs along the propagation direction ($z$-axis). Figure 2C plots the dispersion relation for SPPs at silver-air interface. The dielectric constant $\varepsilon_m$ of silver is a frequency-dependent function given by Drude model, $\varepsilon_m(\omega) = 1 - \frac{\omega_p^2}{\omega(\omega + i\gamma)}$, where $\omega_p$ is the bulk plasmon frequency and $\gamma$ is the damping frequency. Below the surface plasmon resonance frequency $\omega_{sp}$ (at which $\text{Re}(\varepsilon_m) = -\varepsilon_d$), the dispersion curve of SPPs (green solid line) always lies to the right of the dispersion curve of light in the dielectric medium, $k = \sqrt{\varepsilon_d}\omega/c$, the so-called dielectric light line (red dashed line). The large SPP wave vector results in a small SPP wavelength, in comparison with the propagating light in the dielectric medium at the same frequency. The wave vector along the surface normal direction ($x$-axis) is imaginary, implying SPPs are confined at the interface. Above the surface plasmon resonance frequency, SPPs become lossy and quasi-bound. By further reducing the geometric dimensions, the imposed boundary conditions confine SPPs into two dimensions (metallic nanowires) or three dimensions (metallic nanoparticles) at truly nanoscale beyond the Abbe diffraction limit (Figures 2D and 2E). Such localized SPPs are sensitive to the material property, size and shape of the metallic nanostructures. The unique properties of SPPs, that is, subwavelength confinement and strong field enhancement, promise a variety of novel applications in biomedical sensing,[22] super-resolution imaging,[23] energy harvesting,[24] nano manufacturing[25] and next-generation optical circuits.[26] For instance, owing to the excitation of SPPs, resolution of the so-called "superlens" has reached as high as 30 nm,[27] showing the potential for implementing optical microscopy to break the diffraction limit. On the other hand, in plasmonics-assisted spectroscopy, such as surface-enhanced Raman spectroscopy (SERS), the resolution can be even higher. By utilizing the strong local field enhancement of plasmonic nanostructures and the chemical mechanism involving charge transfer between the species and the metal surface, SERS can considerably amplify the Raman signal so that the vibrational information of single molecules can be probed and recorded.[28-30] Tip-enhanced Raman spectroscopy (TERS) further improves this resolution. Very recently, there has been a report on TERS-based spectral imaging with unprecedented sub-molecular spatial resolution (below one nanometer), allowing us to unambiguously resolve the inner structure and surface configuration of a single molecule.[31]

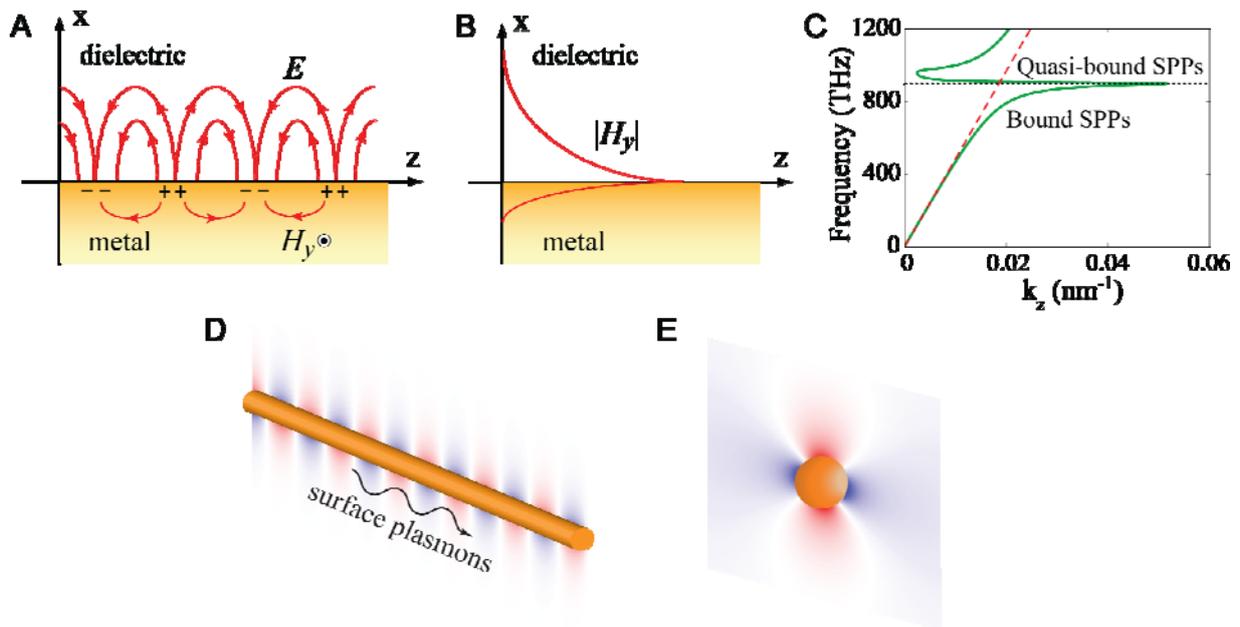

**Figure 2** (A) Illustration of SPPs and oscillation of surface charges at the interface between a dielectric and a metal. (B) The electromagnetic field is maximum at the interface and exponentially decays in the direction perpendicular to the interface, reflecting the bound, non-radiant nature of SPPs. (C) Dispersion curve (green) for SPPs at silver-air interface. The red dashed line is the light line in air, and the black dotted line represents the surface plasmon resonant frequency at which $\mathrm{Re}(\varepsilon_m) = -\varepsilon_d$. (D) and (E) Schematics of SPPs supported by a metallic nanowire and a metallic nanoparticle, which are confined in two dimensions and three dimensions, respectively. The false colour indicates the electric field of SPPs.

The fundamentals, developments and applications of metamaterials and plasmonics have been discussed in many review articles. However, they focus on either metamaterials or plasmonics, rather than the remarkable properties and effects originating from the interplay between these two areas. As demonstrated by a number of groups, combining metamaterials and plasmonics could open up a broad range of opportunities to advance both fields and stimulate new cross-disciplinary approaches. Therefore, it is time to review the recent progress in plasmonic metamaterials. It is beneficial to reflect how the concepts of metamaterials inspire plasmonics and vice versa, potentially giving rise to more breakthroughs in both areas, and in nanotechnology in a broader context.

The rest of the review is organized as follows. We will first discuss different schemes to realize negative refractive index and negative refraction at optical frequencies, followed by the review of recently developed metasurfaces that allow us to generalize the refraction and reflection laws by controlling the phase front. Subsequently we will present THz plasmonics and metamaterials based on graphene, which exhibit extraordinary tunability via electrical gating. One important application of plasmonic metamaterials is biomedical sensing, which will be presented in section 5. We will then discuss some self-assembly techniques to implement sophisticated plasmonic metamaterials with fine features, and finally provide a conclusion and a brief perspective on this fascinating area of plasmonic metamaterials.

## 2. Plasmonic Metamaterials to Implement Negative Refraction and Negative Refractive Index

Negative refractive index and negative refraction are the primary incentive that kindled the research of metamaterials. In 1968, Victor G. Veselago, a Russian physicist, published a theoretical paper studying a material with a negative refractive index.[32] He showed that if the relative electric permittivity ($\varepsilon_r$) and relative magnetic permeability ($\mu_r$) are simultaneously negative, the refractive index of the material has to be negative as well, that is,

$$n = -\sqrt{|\varepsilon_r||\mu_r|} \tag{2}$$

Such a negative-index material (NIM) does not exist in nature, but it exhibits many remarkable properties. For instance, it can be rigorously proved from Maxwell's equations that for a plane wave propagating in a NIM, the wave vector $\vec{k}$, electric field vector $\vec{E}$ and magnetic field vector $\vec{H}$ follow the left-handed rule; and the wave vector $\vec{k}$ and Poynting vector $\vec{S}$ (defined as $\vec{S} = \vec{E} \times \vec{H}$) are antiparallel to each other. This is in a sharp contrast to a positive-index material, in which $\vec{k}$, $\vec{E}$, and $\vec{H}$ form a right-handed triplet, and $\vec{k}$ and $\vec{S}$ are parallel to each other.

The unusual electromagnetic properties of NIMs prompt us to revisit some fundamental phenomena. For example, it is found that Snell's law, Doppler effect, Cherenkov effect and radiation pressure are all reversed in NIMs.[32] Let us focus on Snell's law, one of the basic laws in electromagnetics that determines how light refracts at the interface between two media. As illustrated in Figure 3A, if light is incident from a positive-index material to a negative-index one, the refracted light lies on the same side as the incident light with respect to the surface normal. This is because the tangential components of the wave vectors for incident and refracted light have to be conserved, and the energy must flow away from the interface (causality principle). From Snell's law

$$\frac{\sin\theta_i}{\sin\theta_r} = \frac{n_r}{n_i}, \tag{3}$$

the angle of refraction is indeed negative when the refractive indices of the two materials are opposite in signs. In other words, negative refraction takes place.

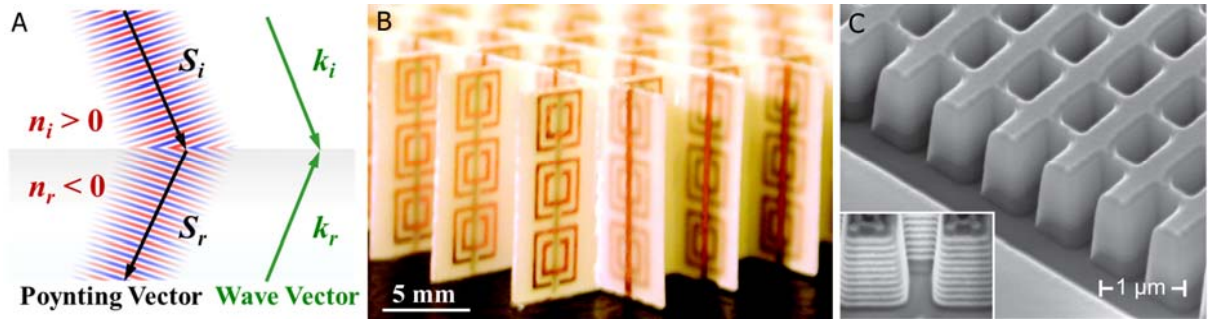

**Figure 3** (A) Illustration of negative refraction when light is incident from a positive-index material to a negative-index material. Note in the negative-index material, the wave vector ($\vec{k}$) and Poynting vector ($\vec{S}$) are antiparallel. The wave patterns (in blue and red) are from a full-wave simulation. (B) Photograph of a negative-index metamaterial working at microwave frequencies, which consists of copper wires and SRRs. (C) Scanning electron micrograph of a 3D fishnet structure to realize bulk NIMs at optical frequencies. The inset shows alternating layers of 30 nm silver (Ag) and 50 nm magnesium fluoride ($MgF_2$). (B) and (C) are adopted from Ref. 33 and Ref. 35 respectively with permissions.

With arrays of copper wires and split ring resonators to implement simultaneously negative permittivity and permeability (Figure 3B), negative index materials and negative refraction were first demonstrated experimentally at microwave frequencies.[8,33] This breakthrough sparked a global interest in metamaterial research. Tremendous effort has been devoted to scale down the size of the building blocks and optimize the metamaterial design to enable operation at higher frequencies. In order to work at optical frequencies, the feature size of metamaterials needs to be sub-100 nanometers or even smaller, imposing a considerable fabrication difficulty. So far, the most successful optical NIMs are the fishnet structure,[34] which significantly eases the fabrication burden. The fishnet metamaterials consist of two layers of metal meshes separated by a dielectric spacer layer. The paired stripes oriented parallel to the electric field provide negative permittivity, while the other pairs of stripes parallel to the magnetic field offer negative permeability. Such a configuration is much simpler than NIMs by combining metallic wires and split ring resonators. By stacking 11 functional layers of fishnet structures (Figure 3C), Xiang Zhang's group demonstrated negative refraction through broadband, bulk and low-loss NIMs at near-infrared wavelengths.[35] Subsequently, loss-free and active optical NIMs in the visible wavelength range are achieved by incorporating gain materials into the fishnet structure.[36]

Engineering the dispersion of SPPs provides an alternative approach to realize negative refraction and NIMs. For instance, there are two bound-SPP modes in a two-dimensional metal-dielectric-metal (MDM) slab waveguide, which originate from the coupling and hybridization of the SPP mode at each individual metal-dielectric interface when the thickness of the dielectric layer is small.[37] Interestingly, for the high-frequency, antisymmetric SPP mode (between the surface plasmon resonance frequency and the bulk plasmon frequency), the dispersion curve exhibits negative slope, leading to negative group velocity ($v_g = d\omega/dk$). In contrast, the phase velocity ($v_p = \omega/k$) is positive. As a result, the energy and phase fronts propagate in opposite directions; this mode thus behaves as if it has a negative refractive index. On the other hand, the low-frequency, symmetric SPP mode (below the surface plasmon resonance frequency) appears to have a positive refractive index, since its group and phase velocities are simultaneously positive. Utilizing these properties, Lezec *et al.* directly demonstrated negative refraction at visible wavelengths by cascading two MDM waveguides.[38] The Au-Si$_3$N$_4$-Ag waveguide is designed to sustain a single negative-index mode, while the Ag-Si$_3$N$_4$-Ag waveguide is tailored to support propagation of only positive index modes within the same wavelength range (Figure 4A). The left column of Figure 4B shows the scanning electron microscope (SEM) image of the fabricated structure to demonstrate in-plane negative refraction, which was made by applying a sequence of thermal evaporation and focused ion milling steps to both sides of a suspended Si$_3$N$_4$ membrane. The input slit at the backside is revealed by electron transparency. In the prism region, the SPPs exhibit negative group velocity for the thickness of the Si$_3$N$_4$ equal to 50 nm, while the rest region only sustains SPPs with positive group velocity. The semi-circular slit works as an output coupler. At $\lambda_0$ = 685 nm, the position of the output spot indicates positive refraction at the prism's interface with a refraction angle θ = +0.1°. The output-spot position shifts upwards when $\lambda_0$ is reduced to 514 nm, indicating negative refraction of SPPs with a refraction angle θ = –43.7°. Given the incident angle equal to 7°, Snell's law yields an effective index ratio for SPPs, which is +0.01 and –5.57 at $\lambda_0$ = 685 nm and 514 nm, respectively.

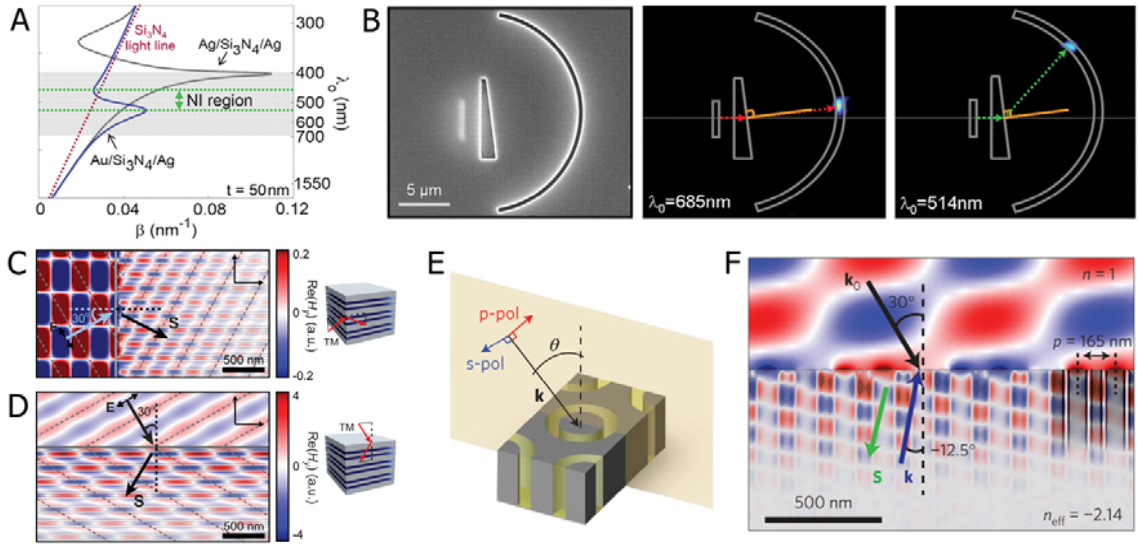

**Figure 4** (A) Dispersion curves for SPPs in an Ag-Si$_3$N$_4$-Ag and an Au-Si$_3$N$_4$-Ag waveguide when the thickness of Si$_3$N$_4$ is 50 nm. (B) Left column: Scanning electron micrograph image of the fabricated device at the output side. Central and right columns: optical microscope image for illumination at $\lambda_0 = 685$ nm and 514 nm, showing positive refraction and negative refraction of SPPs, respectively. (C) and (D) Field plot of the $H_y$ component showing negative refraction of light ($\lambda_0 = 400$ nm), which is incident from air into a 3D metamaterial made of an array of MDM waveguides. The arrows show the unit-cell-averaged direction of the time-averaged Poynting vector. The illuminated interface is normal and parallel to the waveguides in Figure C and D, respectively. Gray dashed lines indicate incident and refracted wave fronts. (E) Unit cell of a NIM consisting of a hexagonal array of subwavelength coaxial waveguide structures. (F) Negative refraction of light ($\lambda_0 = 483$ nm) at the interface between air and a metamaterial made of an array of Ag-GaP-Ag coaxial plasmonic waveguides. The period of the waveguide array is 165 nm, the thickness of the GaP annular channel is 25 nm, and the inner Ag diameter is 75 nm. Shown is a snapshot of the $H_y$ component taken in the polarization plane. (A) and (B) are adopted from Ref. 38, (C) and (D) are adopted from Ref. 39, and (E) and (F) are adopted from Ref. 43 with permissions.

By stacking an array of MDM waveguides, Verhagen *et al.* theoretically demonstrated a metamaterial that have a three-dimensional (3D) negative refractive index at visible frequencies.[39] In the design, the unit cell is a pair of strongly coupled MDM waveguides, in which eigenmodes resemble even and odd superpositions of the antisymmetric mode in the individual MDM waveguide. The odd superposition exhibits a negative mode wave vector as well as a negative coupling coefficient, resulting in negative refraction of both phase and energy for the transverse magnetic (TM) polarized light. The constant frequency contour can be almost spherical by tailoring the dielectric layer thickness between the unit cells. Therefore, wave propagation in such a 3D metamaterial can be described by an isotropic negative refractive index. Figures 4C and 4D show the $H_y$ component when a TM-polarized plane wave ($\lambda_0$ = 400 nm) is incident from air into the metamaterial under two different conditions, that is, from the side and top of the waveguides, respectively. The arrows indicate the calculated time-averaged Poynting vector $\bar{S}$, while the gray dashed lines indicate the incident and refracted wave fronts. From the two figures, one can clearly see that not only the wave fronts, but also energy is refracted negatively with n = –1 for light incident on both interfaces. The negative refraction and lensing effect was experimentally demonstrated in stacked MDM waveguide arrays for ultraviolet light.[40] Very recently, the transformation optics approach is used to conformally map the plasmonic waveguide into a crescent-shaped plasmonic resonator to degenerate electric and magnetic dipoles.[41] Numerical simulations show that a periodic array of such resonators exhibits negative refractive indices over 200 nm wavelength bandwidth at optical frequencies. In addition to negative refractive index, near-zero refractive index has also been demonstrated in the planar MDM waveguide[42] and the transformed crescent plasmonic structure[41] in the optical domain.

The negative refraction in Refs. 38 and 39 only works under TM polarization. To realize polarization-independent or even isotropic NIMs, different approaches have been proposed. For example, similar to the planar MDM plasmonic waveguide, a coaxial MDM waveguide geometry can also support an antisymmetric SPP mode that has negative refractive index (Figure 4E).[43] Thanks to the cylindrical symmetry, the negative-index mode is accessible from free space, independent on polarization and incident angle. It is shown that an array of coupled plasmonic coaxial waveguides functions as a NIM working in the 450-500 nm spectral range with a figure of merit FOM = |Re(n)/Im(n)| ~ 8. Figure 4F plots the $H_y$ component for a TM polarized light ($\lambda_0$ = 483 nm) incident at 30°. The phase front is clearly refracted in the negative direction with an angle of –12.5° with respect to the interface normal, and the wave indeed propagates backward from the finite-difference time-domain (FDTD) simulation. Negative refraction based on arrays of metallic nanowires was experimentally demonstrated in 2008,[44] while the mechanism is the negative group index arising from the hyperbolic dispersion of such a metamaterial.[45,46]

In addition, various attempts have been made to achieve isotropic NIMs by utilizing the Mie resonance of core-shell nanostructures. From the Mie theory, a particle can support both electric and magnetic multipolar modes, and the electromagnetic scattering and extinction efficiencies of a particle originate from the contribution from all the modes. The magnetic dipole resonance is generally much weaker than the fundamental electric dipole mode, but it can be significantly enhanced in a structure made of high-permittivity materials,[47-49] which include semiconductors such as silicon and germanium. Some properly designed core-shell structures exhibit negative refractive indices by overlapping the electric and magnetic resonance within the same spectral range. The metallic core provides the electric plasmon response, while the semiconductor shell is responsible for the strong magnetic resonance with negative permeability. Numerical simulations have shown that both core-shell nanowires and core-shell nanoparticles can implement NIMs operating in the near-infrared regime.[50-52] Because both electric and magnetic responses are obtained from a single constituent, no special requirement is needed for the lattice arrangement of inclusions. In particular, NIMs constructed by core-shell nanoparticles exhibit isotropic and polarization-independent properties, owing to the spherical symmetry of the constituents. Experimental demonstration of such NIMs has not been achieved yet, although strong magnetic response has been observed in a complex plasmonic core-shell geometry.[53]

At the end of this section, let us briefly discuss the reversal of Cherenkov effect, Doppler effect and radiation pressure in NIMs.[32] Cherenkov effect is the electromagnetic radiation of a charged particle when it travels through a medium faster than the phase velocity of light in that medium.[54] In conventional materials with positive refractive index, the radiation falls into an expanding cone in the forward direction of movement, whereas in NIMs the direction of radiation is reversed to the backward side.[32,55] Similarly, inverse Doppler effect occurs when observing a source moving in an NIM. The experimental observation of the two effects is more challenging than that of negative refraction. For the reversed Cherenkov radiation, its power reaches maximum in the ultraviolet region and drops rapidly with increasing wavelengths.[56] The currently available optical metamaterials are still quite lossy, not suitable for detecting the weak radiation signal. Even at microwave regime where both the fabrication and detection techniques are well-developed, only until recently have researchers reported the first successful observation of the reversed Cherenkov radiation,[57] though some groups had previously presented analogous[58] and indirect evidence.[59] For the inverse Doppler effect, the main difficulty originates from the measurement of frequency shift inside the media. So far there has been a few experimental demonstrations of inverse Doppler effect based on nonlinear transmission lines at THz,[60] and photonic crystals with an effective negative refractive index at optical frequencies.[61] However, the demonstration of inverse Doppler effect by means of NIMs is still lacking.

Another intriguing while subtle subject related to NIMs is the negative radiation pressure or negative optical force. According to the prediction in Veselago's pioneering paper, a light beam propagating in a NIM and impinging on a reflecting object exerts a momentum directed backward to the source,[32] unlike in a conventional medium where the resulting momentum points to the direction leaving the source.[62] Therefore, light pressure on the object embedded in a NIM counterintuitively causes an attraction, or a pulling force. However, because the century-old debate on the definition of momentum of light in a dielectric medium,[63-65] namely the Abraham-Minkowski controversy, has not yet been settled, we need to pay special attention when dealing with the radiation pressure or optical force in NIMs.[66,67] Some authors suggest that for NIMs, positive admittance and group refractive index, instead of negative phase refractive index, should be used as material parameters when radiation pressure is concerned. Under such a consideration, the radiation pressure in NIMs is not different from that in conventional materials.[68] The experimental confirmation on the direction of radiation pressure is even more difficult than that of inverse Doppler effect, owing to the similar material property and detection issues. The force of light is normally very weak thus needs to be measured via the motion of small objects. Unfortunately, current optical metamaterials are solids which prevent the movement of any objects.[69] Although the optical force verification in NIMs is intrinsically challenging, research on negative optical forces has recently achieved great progress by other judicious strategies, such as tailoring the incident beam profile,[70,71] the object morphology,[70] and the dielectric environment.[72] All these experiments are performed based on materials with positive refractive index, which exceed the scope of this paper. For a more detailed discussion on negative optical forces, we refer the readers to a latest review article.[73]

## 3.   Plasmonic Metasurfaces

Since 2011, there has been an extensive interest in metasurfaces, a new kind of 2D metamaterials.[74-85] By patterning plasmonic nanostructures and engineering the spatial phase distribution within the plane, we can achieve exotic optical phenomena and optical components, including negative refraction or reflection, as well as ultra-thin focusing or diverging lenses. These findings are remarkable, since normally a material with a finite thickness (with respect to the wavelength) is required to accumulate sufficient phase to shape the beam profile. This requirement applies to classical optical components (such as glass lenses, prisms and wave plates) and also metamaterials (such as a NIM slab for imaging). In comparison, metasurfaces tailor the in-plane phase front using an extremely thin slab consisting of judiciously designed plasmonic structures. To some extent, metasurface can be considered as the optical counterpart of frequency selective surface (FSS),[86] a 2D planar structure widely used in radio-wave and microwave regions under the long wavelength limit. In FSS, each unit cell acts as a small antenna, which is excited by external incidence and then emits or scatters waves with a phase variation relative to its neighbouring units. As a result, specifically tailored wave fronts can be realized. Depending on the characteristic of the scattered waves, an FSS can be used as a lens (transmitarray),[87,88] a reflector (reflectarray),[89] or an absorber,[90] etc. Thanks to the advances in nanotechnology, optical nanoantennas with a feature size below 100 nanometers can be routinely fabricated and precisely arranged. By rationally designing the structural features, one can build optical nanoantenna arrays analogous to FSS. These compact arrays introduce prescribed phase distribution within deep subwavelength thickness, leading to exceptional changes of light propagation characteristics. Here we concentrate on the current development of nanoantenna-based metasurfaces for controlling propagating waves and surface plasmons, while the application of metasurfaces for biosensing will be discussed in section 5.

In conventional optical components consisting of dielectrics, the behavior of light is governed by the gradual phase change accumulated along the optical path, implying that the surface geometry and refractive index profile are the two most important factors. On the other hand, it is well known that phased arrays at radio frequencies, which utilize a group of antennas fed by spatially varied signals to control the radiation pattern, can introduce abrupt phase shifts along the optical path to steer electromagnetic waves. Similar active devices at optical wavelengths have long been investigated aiming at large-scale on-chip integration. A nanophotonic phased array was successfully fabricated recently.[91] For passive devices, as long as the change of phase occurs within a very small thickness, the effect of wave propagation in this layer can be neglected, making the phase discontinuity an inherent characteristic of the interface. Under this circumstance, the standard Snell's law that characterizes refraction and reflection should be rewritten into a generalized form:[74]

$$\sin(\theta_t)n_t - \sin(\theta_i)n_i = \lambda_0 \nabla\Phi / 2\pi, \quad (4)$$

$$\sin(\theta_r) - \sin(\theta_i) = \lambda_0 \nabla\Phi / 2\pi n_i, \quad (5)$$

where $\theta_i$, $\theta_t$, and $\theta_r$ are the angles of incidence, refraction, and reflection, respectively; $n_i$ and $n_t$ are the refractive indices of the two media on the incidence and transmission side respectively; $\lambda_0$ denotes the wavelength in free space and $\nabla\Phi$ is the gradient of phase along the interface. A non-trivial phase gradient will change the directions of the refracted and reflected light simultaneously. This feature is fundamentally different from bulk materials or even metamaterials, where normally only the direction of the refracted component is changed. In addition, as can be deduced from Eqs. (4) and (5), if the abrupt phase shift term $\nabla\Phi$ is properly selected, not only negative refraction but also negative reflection can take place.

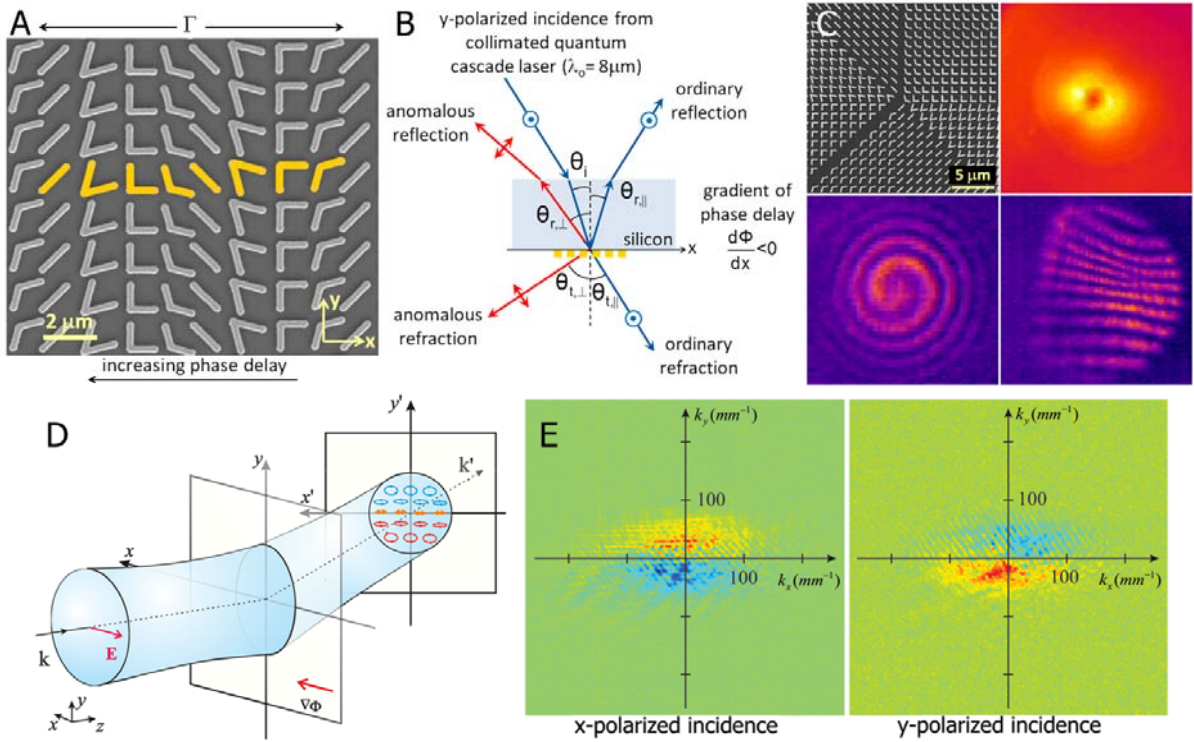

**Figure 5** (A) SEM image of a representative metasurface consisting of a V-shaped plasmonic nanoantenna array. (B) Schematic illustration of metasurface-induced anomalous reflection and refraction, in comparison to the ordinary reflection and refraction. (C) Properly engineered metasurface creates an optical vortex with a topological charge of one. Starting from the top left, the four images in the clockwise direction show the SEM image of the sample, the far-field intensity distribution of the vortex beam, the fringe pattern created by the interference of the vortex and a Gaussian beam tilted with respect to it, and the spiral pattern created by the interference of the vortex and a co-propagating Gaussian beam. (D) Transverse polarization splitting from a metasurface with a phase retardation along +*x* direction and (E) its experimental evidence. For data in (E), red and blue indicate the right and left circular polarizations, respectively. (A-C) are adopted from Ref. 74, and (D) and (E) are adopted from Ref. 78 with permissions.

The physical realization of an interface with continuous phase shifts is not trivial. Fortunately, it can be constructed by an array of plasmonic nanostructures with varied geometries and subwavelength separations. Similar to the long-wavelength counterparts of transmitarrays and reflectarrays in FSS, here the spatially varied nanostructures play the role of resonators to engineer the in-plane phase shifts, which eventually determine the directions of the refracted and reflected beams. Figure 5A shows the SEM micrograph of a metasurface consisting of gold nanoantennas with a V shape.[74] By breaking the geometrical symmetry, a V-shaped antenna gives different responses to the different polarization components of the incident light, providing a wide range of degrees of freedom to modify the phase distribution and the state of the scattered light at the interface. As shown in Figure 5A, the V-shaped nanoantennas are arranged periodically along *x* direction (period 11 μm), and the angle between the axis of symmetry of each unit cell and the *x*-/*y*-axis is fixed to be 45°. For an *x*- or *y*-polarized incidence, these antennas can provide cross-polarized scattered light, with equal amplitudes and constant phase difference between neighbors along the *x*-axis. For the scattered light, the cross-polarized component exhibits anomalous reflection and refraction due to the phase discontinuity across the interface, while the other component simply follows the standard laws, resulting in ordinary propagation. Figure 5B plots the schematic of an experimental setup with *y*-polarized incidence. This 2D demonstration of anomalous refraction and reflection was later extended to the 3D version.[92]

A novel application of metasurfaces is far-field beam shaping.[84,93-99] Because phase shifts are controlled by the arrangement of plasmonic nanoantennas, the wave front of the scattered light can be shaped in an almost arbitrary manner. For example, one can easily generate light beams carrying desirable orbital angular momentum (OAM), so called optical vortices or vortex beams.[74] A vortex beam is an optical field propagating with its wave front twisted around the beam axis like a corkscrew. This unique property promises direct applications in detection[100] and optical tweezers[101,102] for rotating or trapping small particles. The vortex beams have infinite states characterized by an integer called topological charge, which represents the number of twists of the wave front as light propagates a distance of one wavelength. Because different topological charges correspond to different speeds of information manipulation, vortices are expected to be applied in signal processing such as encoding and multiplexing.[103,104] The creation of vortex beams is commonly achieved by using spiral phase plates, computer generated holograms, q-plates or other modulators, which are bulky and rely on the surface geometry and refractive index profile. In contrast, metasurface can easily generate such beams with a proper phase distribution by patterning nanoantenna arrays. The top left panel of Figure 5C shows an example of the fabricated plasmonic metasurface to create vortex beams at mid-infrared with a topological charge of one. The array is divided into eight sectors occupied by one constituent element each, rendering in turn a $\pi/4$ phase delay in the clockwise direction. Using such a structure, vortex beams can be excited upon normal illumination of linearly polarized light. The resulting beam has a uniform intensity except the singularity at the center, which appears as a bright ring in the top right panel. The helical wave front of the vortex is revealed by its interference with co-propagating Gaussian beams parallel and tilted with respect to it, , as shown by the spiral and fringe patterns at the bottom of Figure 5C, respectively.

In addition to OAM, the spin angular momentum (SAM) is another fundamental property of light. In a circularly polarized light beam, each photon carries a SAM of $\pm\hbar$ directed along the beam axis.[105,106] Here $\hbar$ is the reduced Planck constant and the positive (negative) sign is adopted for left (right) circular polarization. When the light beam is propagating along a curved trajectory, the evolutions of polarization state (spin) and light trajectory (orbital motion) will affect each other.[107] This optical spin-orbit interaction, in analogy to its electronic counterpart, is normally weak, requiring the measurement sensitivity of displacements at the angstrom level.[108-110] A rapid phase gradient along a metasurface can dramatically enhance the optical spin-orbit interactions when photons pass through the metasurface,[78] giving rise to clearly observable changes to the trajectories of light besides the anomalous refraction along the phase gradient. This is the so-called helicity-dependent transverse displacement, or photonic Spin Hall Effect (SHE). For an incidence of linear polarization, its components of opposite helicities will be split and accumulated at the opposite half spaces of the anomalously refracted beam, as schematically depicted in Figure 5D. Experimentally, the helicity of the anomalously refracted light cannot be observed directly. However, it can be retrieved for visualization from the measured data.

Specifically, the circular Stokes parameter is determined by the relative intensities of right and left circular polarization, which can be discriminated by wave plates and polarization analyzers and imaged by cameras. A successful observation of photonic SHE based on this scheme is presented in Figure 5E. This progress may open new possibilities for optical information technology.

Metasurfaces exhibit three unique advantages over bulk metamaterials in controlling light propagation with extraordinary abilities. First, the reduced dimension could also reduce both loss and cost, which arises from Ohmic loss of metals and fabrication complexity, respectively. Second, the designs of nanoantennas are diverse, thus allowing one to adopt different candidates for different applications, where the requirements on wavelength, polarization, and efficiency may vary. For instance, a dipole antenna array is demonstrated to work on the anomalous control of circularly polarized light.[84,94] Last, the planar metasurface can be readily integrated with other on-chip or quasi-planar optical devices. This is highly desirable for miniaturized photonic systems. Unlike conventional lenses with bulk volume and complex surface curvature, a metasurface-based lens could be ultra thin, flat, and aberration-free.[84,93,111] We note that recent progress at long wavelength regime is also impressive. In Ref. 112, a microwave metascreen is designed to tailor the wave front without causing any reflection. This dual-functionality is still challenging for the current metasurfaces based on nanoantennas.[113]

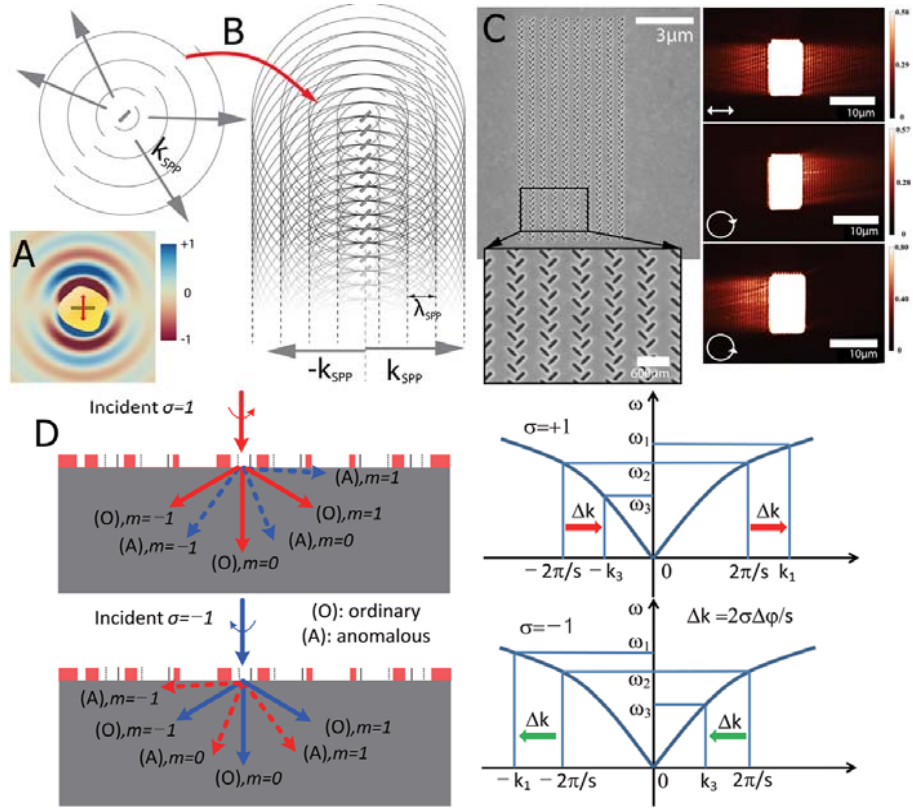

**Figure 6** (A) A narrow aperture in a gold film selectively scatters light polarized perpendicular to its orientation. (B) SPPs emitted from an array of closely-spaced subwavelength apertures interfere constructively along planes parallel to the column. (C) Metasurface consisting of multiple parallel column pairs enables polarization-controlled directional coupling of SPPs. Linearly polarized light generates two SPP beams symmetrically propagating to both sides of the structure; while in contrast, the right (left) circularly polarized light is converted to SPPs propagating to the right (left) side of the structure only. (D) Anomalous diffraction of right (left) circularly polarized light adds an extra momentum towards right (left), leading to asymmetric excitation of SPPs. (A-C) are adopted from Ref. 79, and (D) is adopted from Ref. 81 with permissions.

Another exciting feature of plasmonic metasurfaces, which is revealed very recently, is the ability to control optical surface waves, such as SPPs, in the near-field regime.[76,79-82] Conceptually, one can manipulate the propagation of light and SPPs with the transformation optics approach as they both obey Maxwell's equations.[11,12,114-116] However, the physical realization of transformation optical devices still relies on either bulk metamaterials or dielectrics in most cases, which leads to intrinsic limitations on the size and cost of the devices. The latest work by several groups shows that metasurfaces can efficiently mold SPPs.[79-82] The idea utilizes a fundamental nature of waves, even those bounded at interfaces, that the superposition of two waves may form complex power distribution owing to the constructive and destructive interference. When the patterned plasmonic nanostructures of a metasurface interact with light, part of the re-radiative energy is converted to SPPs. In other words, the plasmonic nanostructures function as SPPs emitters. By controlling the relative phase among the emitters via modifying the geometry of the nanostructures or the polarization state of the incident light, one can achieve unidirectional excitation of SPPs.

The recent work of Capasso's group uses an aperture array—the complementary structure of the nanoantennas, to excite directional SPPs.[79] As shown in Figure 6A, a single subwavelength aperture milled into a gold film resonantly interacts with light that is polarized along the long axis of the aperture, producing an equivalent dipole oriented along the short axis, as represented by the red arrow in the figure. The dipole excites SPPs, which display a symmetric radiation pattern. Then in Figure 6B, the SPPs generated from a column of apertures with the same orientation interfere constructively along the planes parallel to the column, giving rise to plane-wave-like SPPs that propagate symmetrically away from the structure. By placing a second parallel column consisting of apertures oriented orthogonally to those in the first column, the total intensities of the left-going and right-going SPPs can be controlled by the helicity of the incident light and the separation of the two columns of apertures. The left panel of Figure 6C illustrates the SEM image of a metasurface with multiple column pairs to launch polarization-controlled SPPs. The right panels show that a linearly polarized light normally incident from top generates SPP beams of almost equal intensity to both sides of the structure; while in contrast, the right (left) circularly polarized light is converted to SPPs propagating to the right(left) side of the structure only. The different orientations of the apertures between a column pair are essential. The reason is similar to the previous work,[74] in which the V-shaped antennas facilitate the coupling between plasmonic nanostructures and different field components. Near-field interference can be also induced on a flat interface by a circularly polarized dipole, whose vertical and horizontal component owns an odd and even parity of spatial frequency spectrum respectively, resulting in an asymmetric profile of SPPs. As proven in Ref. 80, a single slit obliquely illuminated by circularly polarized light can indeed excite directional SPPs as well.

An alternative route stems from the phase discontinuity mechanism. The metasurface consisting of an aperture array can provide a phase gradient as a dipole antenna array does. As we have discussed, this structure is able to control anomalous light propagation depending on the polarization state. When light illuminates on such an interface, the diffraction equation needs to be modified by taking into account the effect of the phase shifts:[81]

$$\sin(\theta_t)n_t - \sin(\theta_i)n_i = m\lambda_0/s + \sigma \cdot \lambda_0 \nabla\Phi/\pi, \qquad (6)$$

where $\theta_t$ and $m$ are the angle and order of diffraction respectively, $s$ is the lattice constant of the array, and $\sigma$ is a coefficient of helicity taking 1/−1 for the left/right circular polarization. Figure 6D compares the ordinary and anomalous refraction and diffraction of two helicities at normal incidence. While the ordinary diffraction orders lie symmetrically about the surface normal, the anomalous ones are shifted sideward together with the refracted light in accordance with the helicity of the incidence. This asymmetric profile enables one side of the array to gain a larger wave vector, which leads to efficient excitation of SPPs. In contrast, the wave vector is smaller on the other side (shifted closer to the light line), and the SPP excitation condition is not satisfied.

One problem is the coupling efficiency of SPPs. The efficiency achieved in experiments was low, about 2.5% and 3.8% in Ref. 79 and Ref. 81, respectively. This can be improved by optimizing the array configuration and employing highly resonant structures. For instance, a compact magnetic resonator pair can generate directional SPPs with efficiency as high as 135%,[117] enabling better conversion of light from free space to surface plasmons. With these recipes and better plasmonic materials, the coupling efficiency was enhanced up to 10~14% in simulations and is expected to reach a higher level.[81,93]

The family of metasurfaces is actively growing, and there are several directions might be worth exploring in the coming years. The symmetry property of nanoantennas is a possible gateway leading to highly efficient plasmonic sources and spin-controlled photonic devices. Hyperbolic metasurface inherits the superior ability of its 3D counterpart in terms of large photonic density of states.[77,118] Therefore, they are promising for on-chip lasing and quantum information applications. Other issues, like tunability, loss, and so forth, are also important aspects that always exist in plasmonic metamaterials.

## 4. Plasmonics and Metamaterials Based on Graphene

Graphene, a monolayer of carbon atoms arranged in a honeycomb lattice,[119,120] has attracted a great deal of attention for plasmonics and metamaterials at terahertz and infrared frequencies.[121-126] The dispersion of conventional SPPs supported by noble metals approaches the light line in the long wavelength regime, and the subwavelength confinement is completely lost. However, graphene can support surface plasmons to confine light strongly while potentially maintaining small losses in this frequency range because of its unique electronic property.[123,127] Moreover, the mechanical, electronic, optical, and even thermal properties of graphene are highly tunable via chemical doping and electrical gating.[128-134] These favorable features arouse enormous interests on the investigation of graphene-based tunable plasmonics and metamaterials.

Theoretically, surface plasmons can exist in a material with mobile charge carriers whose conductivity is predominantly imaginary.[135] This is equivalent to a material, such as a noble metal, of which the real part of electric permittivity is negative. Graphene is an atomically thin semiconductor, whose electron density, and thus the plasma frequency is much lower than that of bulk noble metals. As a result, graphene satisfies the requirements of supporting SPPs as a bulk metal does, whereas its operating frequencies reside at terahertz and infrared regimes, much lower than that of metals at visible and ultraviolet. Because of the 2D nature of the platform, graphene surface plasmons exhibit exponential field decay into the surrounding dielectric layers on both sides of the graphene sheet (Figure 7A).[123] Indeed, graphene plasmonics resembles plasmons in the two-dimensional electron gases (2DEGs) that have been explored before.[136-138]

As a zero-gap semiconductor, graphene can be doped to high density of charge carriers by either chemical methods or varying the biased gate voltage. The latter approach, known as electrical gating, is particularly preferred as it provides an *in situ* and effective access to tune the electronic and optical properties of graphene layers.[139-141] Figure 7B illustrates one valley of the electronic band structure of graphene, where the intra- and inter-band transitions are represented by the green and pink arrows, respectively. A sufficiently high level of doping leads to a larger value of Fermi energy $E_F$. The absorption of photons with energy less than $2E_F$ is forbidden due to the Pauli blocking. Therefore, doping can suppress the absorption losses introduced by the inter-band transition and compels SPPs towards higher frequencies.[121,124]

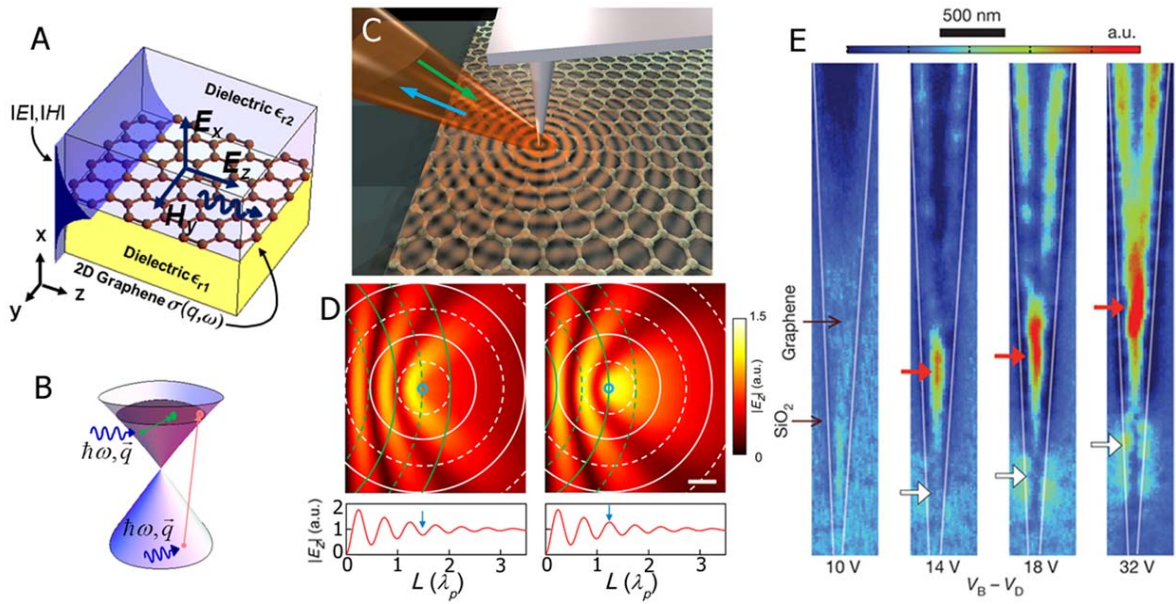

**Figure 7** (A) Schematic of SPPs in graphene. (B) Electronic band structure of graphene. The green and pink arrows represent the intra- and inter-band transitions, respectively. (C) Schematic of the experimental configuration to launch and detect SPPs on graphene based on near-field scanning optical microscopy (NSOM). A sharp metallic tip illuminated by an infrared laser is used to locally excite SPPs, and the near-field SPP signal can be scattered by the tip into the far-field and detected. (D) Top panels show the snapshots of destructive (left) and constructive (right) interference of plasmon waves underneath the tip (blue circles). The bottom panels are the profiles of electric field amplitude underneath the tip versus its distance to the left edge. The scale bar is half wavelength of SPPs. (E) Near-field amplitude images showing the switching of graphene plasmons when a backgate voltage is applied. Red and white arrows denote the positions of resonant local plasmon modes on a tapered graphene ribbon. (A) and (B) are adopted from Ref. 123, (C) and (D) are adopted from Ref. 135, and (E) is adopted from Ref. 145 with permissions.

Because the charge carriers in graphene are massless Dirac electrons and the excitations are essentially confined in the transverse dimension, graphene plasmons are expected to provide much smaller wavelengths and thereby much stronger light-matter interaction. From the dispersion relation, one can calculate the SPPs wavelength on a graphene sheet, which is about 200 nm at infrared frequencies,[127,135] approaching two orders of magnitude smaller than the wavelength of illumination. Such an exceptional confinement is not possible in conventional SPPs supported by noble metals at the same frequency range. However, this feature brings a technical problem: the excitation and detection of graphene plasmons can hardly be efficient due to their large wave vector mismatch with photons in the free space. Although successful observations have been reported via electron spectroscopy[142,143] and tunneling spectroscopy,[144] the spectrally indirect probing is not sufficient to look into the underlying physics of graphene plasmons. Recently, two groups independently reported the direct spatial mapping of graphene plasmons based on the scattering-type near-field scanning optical microscopy (NSOM)[135,145] at infrared frequencies. Thanks to the advanced experimental techniques, high-resolution imaging of the real-space profiles of propagating and standing surface plasmons are achieved. Figure 7C shows the configuration of the experimental setup, which was employed in both studies. Plasmons with high in-plane momenta are excited by illuminating a sharp atomic force microscope (AFM) tip with a focused infrared beam, as indicated by the green arrow. The near-field amplitude profiles are obtained from the back-scattered light beam (blue arrow). The tip is extremely sharp, with the curvature radius of around 25 nm, ensuring excitation of large momentum and high order of spatial resolution. The reported wavelength of graphene plasmons is about 200 nm, which is more than 40-fold smaller than that of the incident infrared light. Various characteristics of SPPs can be observed from the resulting images. For instance, the plasmons interference patterns are given by the simulated near-field amplitude profiles in Figure 7D. Because of the presence of the graphene/substrate boundary at $L = 0$, interference occurs between the tip-launched plasmons (white circles) and the reflected plasmons (green circles), giving rise to different standing wave patterns when the tip is located at different distance from the edge, as denoted by the blue circles and arrows respectively.

A unique property of graphene plasmonics is that the plasmon wavelength is determined by the carrier density, implying one can tune the characteristics of graphene plasmons and achieve sophisticated functionalities by simply varying the gate bias.[145] Figure 7E shows the plasmonic switching by electrical gating, where tapered graphene ribbons are deposited on a $SiO_2$ substrate with a Si backgate. As can be seen, with a sufficiently large and increasing backgate voltage $V_B$, which causes larger carrier density and Fermi energy, the two local plasmon modes shift towards regions with a larger ribbon width due to the corresponding increase in plasmon wavelength. This exciting result clearly manifests the effectiveness of graphene gate-tuning and the capability of graphene plasmonics to control and switch optical signals at the nanoscale.

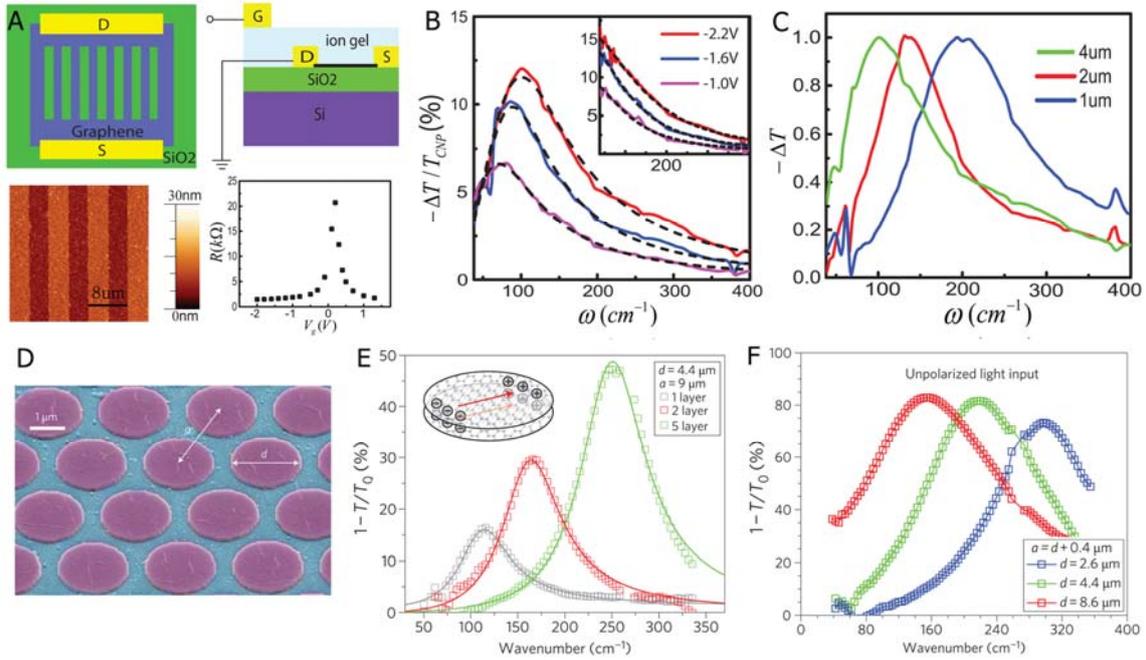

**Figure 8** (A) Graphene micro-ribbon array. Four panels in the clockwise direction show the top-view, side-view, gate-dependent electrical resistance trace, and the AFM image of a sample with the graphene ribbon width of 4 μm and the period of 8 μm. (B) Control of the absorption of graphene plasmons via electrical gating. The graphene ribbon array is the same as in (A), and the polarization is perpendicular to the grapheme ribbon. For comparison, the inset shows the corresponding spectra for terahertz radiation polarized parallel to ribbons. In this case, the absorption strength increases with carrier concentration (bias voltage) but the spectral shape remains the same. (C) Spectra of change of transmission with different micro-ribbon widths (1, 2, and 4 μm). (D) SEM image and (E) extinction spectra of patterned graphene/insulator stacks with varied numbers of graphene layers. (F) Tunable terahertz filters using stacked graphene devices with different diameters of the stack. The total number of graphene layers is fixed at five. (A-C) are adopted from Ref. 139, and (D-F) are adopted from Ref. 141 with permissions.

In addition to doping and gating, geometry patterning is an alternative approach to tune the optical properties of graphene, rendering a variety of graphene-based plasmonic and metamaterial structures at infrared wavelengths. The light-plasmon coupling in pristine graphene is relatively weak and not sufficient to serve practical devices. However, some recent experimental work demonstrated that rationally designed graphene patterns can dramatically enhance the coupling and improve the light-matter interaction.[139,141,146-149] Therefore, patterning is an effective recipe to tune graphene plasmons, which not only diversifies the design of devices[125,150-152] but also enables the investigation of the damping effect in graphene nanostructures.[153] As a representative example, graphene micro-ribbon arrays are investigated in Ref. 139 and show interesting properties. The structure of a gated graphene ribbon array is shown in Figure 8A. The widths of ribbon and gap are identical and varied in experiments from 1 to 4 μm, which is much smaller than the wavelength of incidence and enables efficient excitation of plasmon resonances. Owing to the anisotropic nature of configuration, the micro-ribbon array responds differently to incident light with polarization perpendicular and parallel to the ribbons. In Figure 8B, the spectra of perpendicular polarization exhibit a distinctly different Lorentz line shape, because the plasmon excitations correspond to carrier oscillations across the width of micro-ribbons. While the carrier concentration $n$ increases with increasing gate voltages, the absorption peak not only gains strength but also shifts to a higher energy, in accord with the $n^{1/4}$ scaling signature of massless Dirac electrons (Figure 8B). For comparison, the inset of Figure 8B depicts the absorption spectra for parallel polarization at three different gate voltages. In this case, the optical response originates from mobile charge carrier oscillation similar to that in a graphene monolayer, which can be described by Drude model and changes only in amplitude with increasing gate voltage. A noticeable fact is, in both cases, the absorption at plasmon resonances reaches the level of over 12% for gate voltage greater than 2V. Compared with the 2.3% absorption by a single layer of pristine graphene,[154] such a substantial absorption implies strong light-plasmon couplings in the patterned graphene ribbons.

For polarization perpendicular to the ribbons, changing the width of ribbon is an alternative way to tune the plasmon resonance frequency. Figure 8C reports the change of normalized transmission with different ribbon widths $w$, in which the carrier concentration is fixed to be $1.5 \times 10^{13} \mathrm{cm}^{-2}$. Theoretical derivations based on quasistatic description suggest that the plasmon resonance scales as $w^{-1/2}$, a characteristic behavior of 2DEGs. The experimental results fit well with the prediction as indicated by the blue-shifting trend of resonance with decreasing ribbon widths. Further engineering, such as broadening the ribbon width and adjusting the filling ratio, can be performed to achieve tunable plasmon resonance ranging from 1 to 10 THz.

The strong dependence of polarization may not be a desirable property in practical applications. Graphene/insulator stacks, in a circular shape and arranged in a high-symmetry order, can circumvent such a disadvantage.[155] More importantly, the number of the stacked layers significantly influences the plasmonic resonance frequency and amplitude.[141,156] Figure 8D shows the SEM image of an array of microdisks arranged in a triangular lattice, where two parameters, the diameter of disks $d$ and the lattice constant $a$, determine the configuration of the stack array in the transverse dimension. In Figure 8E, the extinction spectra of the samples are plotted for 1-, 2- and 5-layer samples, respectively. With increasing number of graphene layers, both the resonance frequency and peak intensity increase significantly because of the strong Coulomb interaction between adjacent layers, which is drastically different from the simple addition of carrier concentration in conventional 2DEG lattices. On the other hand, unlike resonances in graphene monolayer owning a weak $n^{1/4}$ carrier density dependence, plasmons in the stack structure show a strongly enhanced dependence on carrier concentration. The resonance amplitude and frequency are proportional to $n$ and $n^{1/2}$, respectively. The latter scale stands for conventional 2DEGs. As a result, one can effectively tune the resonance frequency and amplitude of the graphene plasmons in the stacked microdisks, promising great potential in THz photonic applications. By adjusting the in-plane geometrical parameters while maintaining a large filling factor, which is defined as microdisk area over total area, a peak extinction as high as 85% can be achieved (red curve in Figure 8F). The device functions as a tunable, polarization-independent plasmonic notch filter, and the peak value in the spectra corresponds to an extinction ratio of ~8.2 dB.

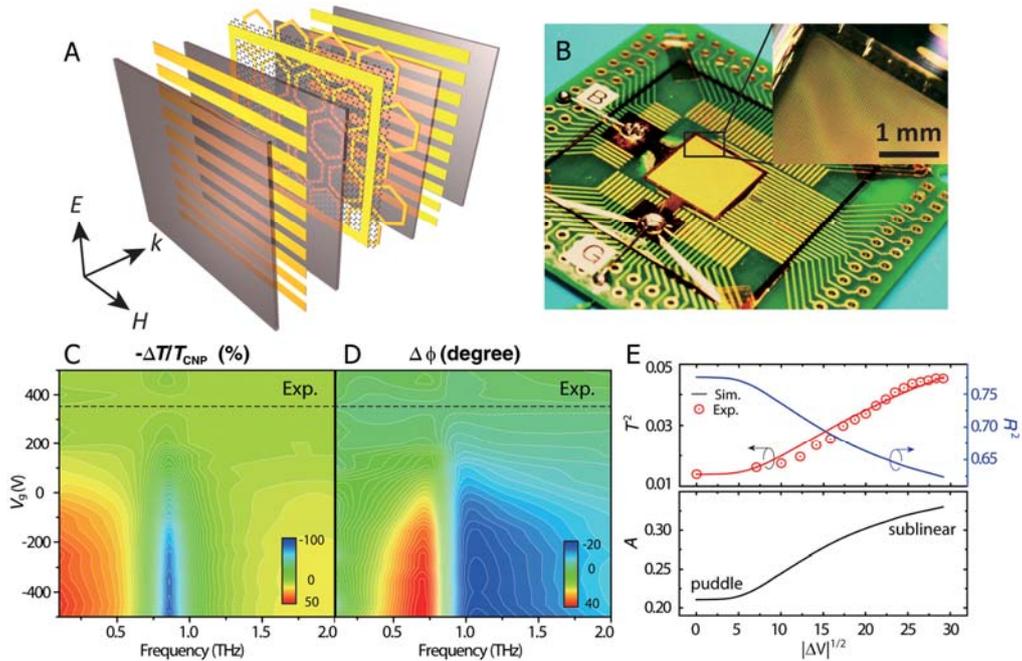

**Figure 9** (A) Schematic of a gate-controlled active graphene metamaterial. A single-layer graphene is deposited on a hexagonal metallic meta-atom layer and connected to an electrode (yellow frame) playing the role of ground, and then sandwiched by EOT electrodes (yellow strips) embedded in dielectrics. (B) The graphene metamaterial in (A) integrated to a drilled PCB for THz-TDS measurement. The inset shows the magnified view of the planar multilayer structure. Measured spectra of (C) relative change in transmission, and (D) phase change of the hexagonal graphene metamaterials. The dashed line denotes the voltage of the charge neutral point (CNP) $V_{CNP}$. (E) The spectra of transmission ($T^2$), reflection ($R^2$), and absorption ($A$) of the graphene metamaterial at the resonance frequency. Figures are adopted from Ref. 140 with permissions.

Because of the extraordinary electronic properties and great tunability, graphene is regarded as a competitive candidate for many applications, such as absorbers,[140] nanoantennas[151,157,158] and waveguides.[125,150,159] To further improve the performance of graphene-based devices, besides manipulation on graphene itself like gating and patterning, the hybridization of graphene and other conventional plasmonic nanostructures or metamaterials is a promising direction. Such scenarios may generate a wide range of tunable optical devices at terahertz and infrared frequencies. For instance, Lee *et al.* recently demonstrated a tunable THz metamaterial by integrating metallic meta-atoms with a gated graphene sheet.[140,160] The schematic plot and optical micrograph of the fully integrated structure are shown in Figures 9A and 9B, respectively. To fabricate the sample, a graphene sheet grown by a chemical vapor deposition (CVD) process is transferred onto a hexagonal meta-atom layer and connected to a grounded electrode. Thin metallic wire arrays are attached to both sides of the hybrid metamaterial as top and bottom electrode, which are designed to provide gate voltage while allowing light to transmit. The polarization of the incident terahertz wave should be perpendicular to the line electrodes in order to induce the extraordinary optical transmission (EOT).[161,162] Here the meta-atom layer plays a key role to enhance the electronic and optical properties of graphene by its strong resonance. The gate-controlled light-matter interaction can be interpreted as follows from two aspects: On one hand, the amplitude of on-resonance transmission increases with increasing gate voltage, as the graphene layer becomes more conductive and suppress the intrinsic resonance of hexagonal meta-atoms. On the other hand, the off-resonance transmission acts in the reverse trend due to the gate-induced broadband absorption in graphene layer, which is dominated by the contribution of intra-band transitions and thus cannot be reduced for larger gate voltage. Terahertz time-domain spectroscopy (THz-TDS)[163] measurement reveals exceptional performance of the device. The amplitude of the transmitted wave is modulated up to 47% at the maximum gate voltage and system resonance frequency, as shown in Figure 9C. This result far exceeds the value of ~30% achieved by metamaterial-semiconductor hybrid structures.[140] The functionality of phase modulation is relatively straightforward considering the resonant nature of the meta-atoms. The maximum phase change is measured to be 32.2° at 0.65 THz (Figure 9D). Further improvement of both amplitude and phase modulation can be achieved when multilayers of graphene, instead of the single sheet, are integrated with metamaterials. The property of such hybridized graphene metamaterials strongly relies on the gate voltage. As seen from Figure 9E, the on-resonance intensity transmission, reflection, and absorption traces exhibit three distinct scaling regimes from left to right, which correspond to the weak, linear, and sublinear dependence of the Fermi level on the voltage $|\Delta V|^{1/2} = |V_{CNP} - V_g|^{1/2}$ (where $V_{CNP}$ and $V_g$ are the voltage of charge neutral point and gate voltage, respectively), because of the conductivity variation of the graphene layer. Interestingly, this active graphene metamaterial also exhibits gate-controlled optical hysteresis that can be applied to develop electrically controlled photonic memories. The design of the active metamaterials could be diverse since the richness of the meta-atoms. In addition to tunable metamaterials, several groups have reported tunable plasmonics by graphene. The gate-dependent dielectric properties of graphene enable the modulation of plasmon resonance frequency and quality factor. Successful demonstrations have been observed on nanoparticles,[164] nanorods,[165,166] and bowtie resonators,[167] etc., all of which are common building blocks to construct plasmonic devices.

The combination of graphene physics and plasmonics generates a versatile new platform for applications in the terahertz and optical regime. By maintaining the advantages of conventional plasmons, graphene-based devices are able to improve the performance of nanoantennas, broadband absorbers, and bio-/chemical sensors; while by keeping the ultra-thin and highly tunable properties, they also have the potential to be used as building blocks of new emerging materials, such as hybrid metamaterials and metasurfaces. Along with the deeper insight of graphene plasmonics, we expect this branch of nanomaterials to bear more groundbreaking results.

## 5. Plasmonic Metamaterials for Biosensing

As discussed in the introduction, SPPs exhibit unique features of tight field confinement and strong field enhancement. They are very sensitive to the changes of refractive index of the dielectric media in the vicinity of metal structures. Taking advantage of these properties, one can design miscellaneous structures for sensing purposes,[22,168-172] through which observable changes of corresponding quantities, such as a shift in resonance,[173-175] a variation in field intensity,[155,176] or an offset in angle,[77,177] are expected. Therefore, SPPs not only provide a label-free, real-time access to probe biological components at nanoscales, but also lead to an extremely high resolution. Both propagating and localized SPPs have been applied for biochemical or biomedical applications. They each have unique advantages while facing their own problems.[173] For approaches based on propagating SPPs, a detection sensitivity exceeding $10^{-5}$ refractive index units (RIU) has been achieved. However, this target is still challenging for analytes with very small size or selective nano-architectures. As to localized SPPs, by altering the size, shape, and chemical components of the plasmonic structures, a device of this kind owns great tunability and adaptability whereas suffers a sensitivity of at least one order of magnitude lower.

Plasmonic metamaterials hold new promise for biosensing with unprecedented sensitivity and specificity. On one hand, it inherits the feature of metamaterials that the operation frequency can be widely tuned to bind the characteristic vibrations of biomolecules, and the spectral line shape can be engineered to be very sharp to obtain higher sensitivity.[176] On the other hand, the plasmonic nanostructures serve as building blocks to keep the merit in adaptability, ensuring that the device satisfies the selective requirements in different applications. In this section, we briefly review recent advances in biosensing based on plasmonic metamaterials in four schemes, *i.e.*, using Fano resonance, nanorod structures, metasurfaces, and chiral fields, respectively. From the limited space, we are not able to discuss other sensing approaches. Interested readers can refer to several latest pieces of work and reviews on these topics.[178-182]

**5.1 Fano Resonance in Plasmonic Metamaterials**

One recent research interest in plasmonic matamaterials is the Fano resonance.[176,182-185] Fano resonance is named after the Italian physicist Ugo Fano, who first suggested the theoretical explanation for the asymmetric scattering line shape of the inelastic scattering of electrons of helium.[186] Only until the recent decade was Fano resonance demonstrated in other systems. The origin of Fano resonance arises from the interference between a discrete narrow state and a wide continuum regardless of the nature of the system. Optical systems consisting of plasmonic nanostructures can be designed to achieve Fano-like spectral responses, which are suitable candidates for sensing, switching,[187,188] and lasing applications,[189] etc.

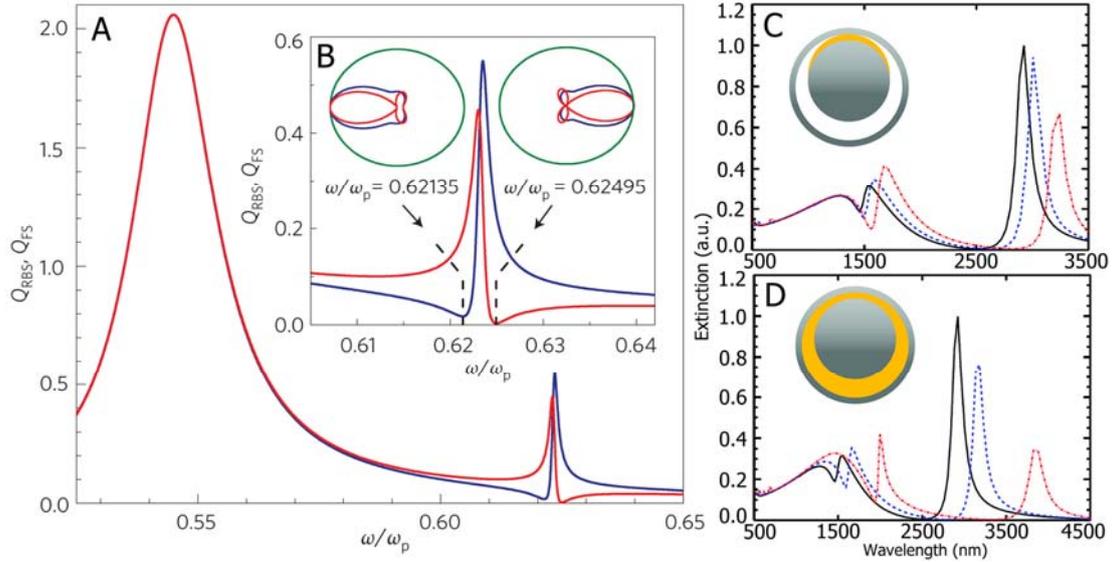

**Figure 10** (A) and (B) Fano resonance in the differential scattering spectra of a plasmonic nanoparticle. (A) Spectra of radar back scattering (RBS) and forward scattering (FS) that exhibit Fano-like asymmetric line shape. The angular dependence of scattering patterns at two frequencies in the vicinity of the quadrupolar resonance changes dramatically, with orientation from backward (red) to forward (blue). In the scattering diagram insets of (B), the closed curves in red and blue represent linearly polarized and non-polarized incidence, respectively. (C) and (D) Fano resonances in nonconcentric ring/disk cavities with (C) partial and (D) complete filling of dielectrics with permittivity 1 (black solid), 1.5 (blue dashed), and 3 (red dotted). (A) and (B) are adopted from Ref. 184, and (C) and (D) are adopted from Ref. 191 with permissions.

Fano resonance exists in a variety of plasmonic structures and metamaterials. For example, it can occur through the light scattering from a metallic nanoparticle.[184,190] According to Mie theory, the scattering amplitudes of a nanoparticle involve multipolar resonant terms for both electric and magnetic responses, which correspond to different localized SPP modes. As the fundamental mode, the electric dipolar resonance is broadband and always the dominant term at low frequencies. However, higher-order modes, originating from the quadrupole, octupole, and so forth, become significant when the frequency increases. These high-order resonances are narrowband, thus satisfying the condition of Fano resonance when they are coupled to the dipolar mode. Figure 10A plots the forward and backward scattering diagrams of a plasmonic nanoparticle, which show distinctly different features for the first two eigenmodes. The radius of the nanoparticle $a$ is much smaller than the wavelength at dipolar resonance $\lambda$ ($a/\lambda = 0.083$), which ensures that the magnetic resonances are negligible and only the electric resonances need to be considered. Constructive and destructive interference occurs near the quadrupolar resonance, rendering the line shape of the spectrum a typical, sharp asymmetric Fano profile. Further connection to biosensing is straightforward.[182] The scattered field intensity in the vicinity of the quadrupolar resonance changes so dramatically that even a slight detuning of frequency will induce astonishing transformation in the scattering patterns, as shown in Figure 10B. This sensitivity is highly desirable in sensing situations to probe the tiny refractive index change.

In coupled nanostructures, Fano resonance is observable from the scattering spectra as long as the size of the system is appropriate. For a single nanoparticle, interference occurs between the "bright" dipole and "dark" higher order multipoles of the particle itself. Herein by contrast, each metal-dielectric interface supports a set of resonances, and both superradiant (bright) and subradiant (dark) states originate from the complex coupling of them, with individual plasmons oscillating in phase and out of phase, respectively. Normally the dipole-dipole coupling dominates the response of symmetric structures such as a nanoshell. However, to achieve Fano resonance for larger field enhancement and better spectral tunability, one needs to break the symmetry exciting higher-order multipolar modes.[23,191-197] A representative example[191] is shown in Figures 10C and 10D, where the extinction spectra for normal incidence are calculated for nonconcentric ring/disk cavities with partial and complete filling of different dielectric insertions, respectively. For each curve, results of charge density calculation reveal that the narrow resonance at long wavelength is due to the anti-symmetric alignment of the dipolar modes in the disk and ring, while the Fano profile around 1500 nm comes from the interaction between the quadrupolar ring mode and the broadband dipolar disk mode. Both spectra exhibit noticeable red-shifts sensitive to the dielectric permittivity increase, and further comparison proves the Fano resonance in the complete filling cavity provides the largest figure of merit for sensing. Such nanocavities are highly tunable in design, involving various planar configurations such as ring/disk dimmers,[191] asymmetric split rings,[193] and dolmen-style slabs.[194] In particular, the latter gives a sharp resonance and is often used in plasmon induced transparency,[23] a plasmonic analogue of the electromagnetically induced transparency (EIT) in atomic

physics.[195,196]

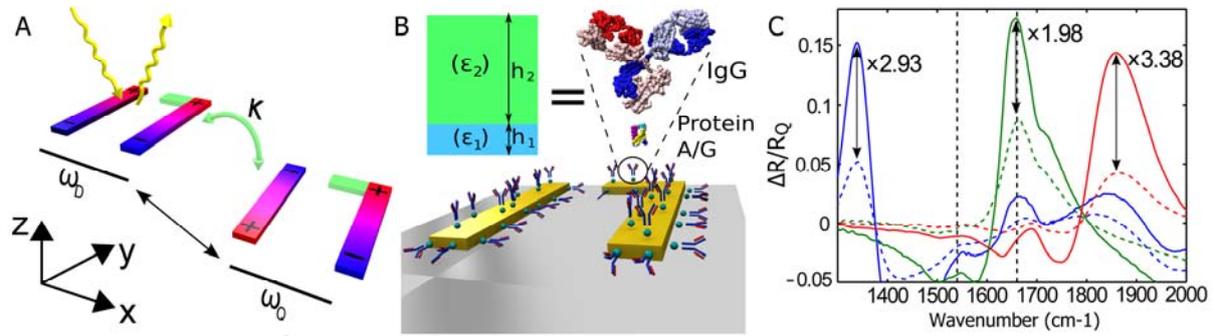

**Figure 11** Biosensing based on Fano-resonant asymmetric metamaterials. (A) Charge density plots of the superradiant ($\omega_D$) and subradiant ($\omega_Q$) modes coupled to infrared incidence. (B) Schematic diagram of proteins binding to metal surface and the equivalent dielectric model. (C) Experimental spectra before (dashed) and after (solid) binding of IgG antibodies to three different FRAMM substrates (indicated by different colors) with immobilized protein A/G. Figures are adopted from Ref. 176 with permission.

Combining the advantages of sharp spectral response, large field enhancement, and tunable frequency of operation at infrared, Fano resonance in plasmonic metamaterials is a promising approach for biosensing applications. For instance, Wu et al. successfully demonstrated Fano-resonant asymmetric metamaterials for multispectral biosensing of nanometer-scale monolayers of recognition proteins and their surface orientation.[176] As shown in Figure 11, similar to the dolmen-style slabs, the FRAMM consists of three nanobar antennas, of which two are parallel and identical, and the third is perpendicularly attached to one of them serving as a symmetry-breaking coupler. The super- and sub-radiant modes contributing to the Fano interference arise from the parallel (in phase) and anti-parallel (out of phase) charge oscillations in the two identical branches (Figure 11A). In measurements, when protein nanolayers are immobilized on the surface of the FRAMM, the optical effect of each monolayer is equivalent to a dielectric film of the same thickness, which induces a spectral shift and intensity difference in the reflectivity spectrum with a unique fingerprint. The topmost protein monolayer can be detected and characterized by measuring the difference of reflectivity between substrates before and after the bonding process. Thereby in principle, a structure-specific biosensing can always be accomplished by using the FRAMM substrate with recognition protein immobilized on its surface. Figure 11B illustrates the sensing concept schematically. A monolayer of protein A/G is immobilized on the surface of the FRAMM substrate as recognition moiety, and IgG antibodies are the target molecules to bond to the protein A/G. Comparisons of the experimental spectra before and after bonding of IgG are shown in Figure 11C with dashed and solid curves respectively. For all the three FRAMMs with different resonant peaks, which are tuned either away from or precisely matching with the eigenfrequency of a biomolecule's vibrational modes, the bonding of IgG changes reflectivity in evidently detectable amplitude. Highly specific information including the thickness and orientation of the bonding biomolecules can be derived with nanometer-scale accuracy from the measured results at different resonances.

**5.2 Nanorod Metamaterials**

In terms of biosensing using localized SPPs, various nanoparticle shapes like spheres, rings, ellipsoids, and rods are adopted. Among these candidates, nanorods, which possess relatively high sensitivity on refractive index changes and geometrical tunability, have been investigated most intensively.[198] Figure 12A shows the basic schematic of a nanorod-based immunoassay.[199] The nanorods are fixed on a glass substrate and coated with self-assembled monolayers (SAMs). The individual nanorod is about 15 nm in diameter and 50 nm long. To perform sensing, capture antibodies are pre-coupled to the SAMs, and specific antigens in the analyte solution will bind to them during the measurement. The extinction peak responds to this reaction with a spectral shift, which is relevant to the concentration of the target analyte and can be monitored via real-time analysis, as shown in Figure 12B. According to the tested results, this design yields a detection sensitivity of 170 nm per RIU and an FOM of 1.3.

Plasmonic metamaterials made of an array of nanorods can introduce new modes to serve as the performer of detection,[200] which are expected to overcome the sensitivity limit of sensors based on localized SPPs and provide better adaptability. Recently, Kabashin *et al.* demonstrated a neat prototype by using an assembly of parallel Au nanorods (Figure 12C),[173] which are electrochemically grown in a porous, anodized alumina template. The structural parameters are highly tunable during fabrication, with rod lengths ranging from 20 to 700 nm, diameters from 10 to 50 nm, and separations from 40 to 70 nm, respectively. From the viewpoint of metamaterials, this subwavelength nanorod array behaves as a bulk material with unusual hyperbolic dispersion relation, whose permittivity tensor contains a negative component along the long axis of nanorods and positive components along the short axis. With this unique feature, although the field distribution in the porous layer is determined by plasmon-mediated interaction between nanorods, the whole structure supports a guided photonic mode at near-infrared. The excitation condition of this guided mode is similar to excite SPPs on a metal film, that is, TM polarization and attenuated total reflection (ATR). For isolated nanorods with a sufficiently large separation distance, the near-field coupling is negligible and usually it is the localized SPPs dominate the response. On the other hand, the guided mode arising from the collective plasmonic response in the nanorod metamaterial is able to reach an extremely high sensitivity of 32000 nm per RIU and a FOM of 330, far exceeding those of sensors based on localized SPPs, and even propagating SPPs.

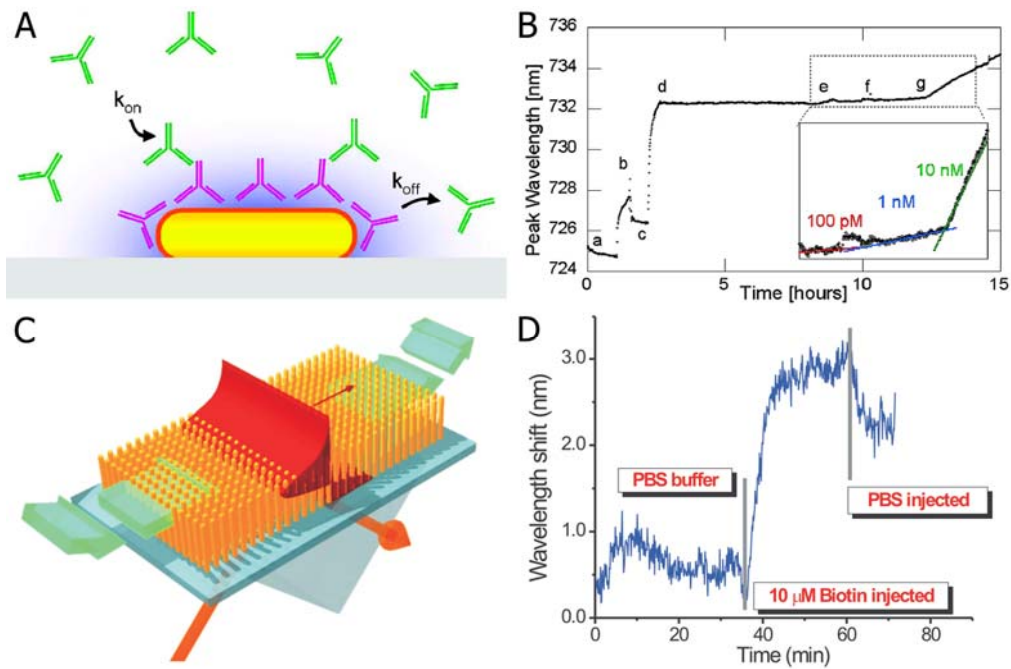

**Figure 12** (A) Nanorod-based immunoassay and (B) the test of sensing sensitivity. The curve in (B) represents the trend of extinction peak drift versus time, with the slope proportional to analyte concentration. The inset shows linear fits to the binding curve when the analyte concentration is raised from 100 pM to 1 nM and then 10 nM. (C) Schematic and (D) real-time response of the plasmonic nanorod metamaterial for sensing experiments. The binding of biotin-streptavidin causes an immediate increase of wavelength shift until its saturation. In comparison, the removal of unbound and weakly attached biotin by injecting PBS buffer leads to a rapid decrease of shift. (A) and (B) are adopted from Ref. 199, and (C) and (D) are adopted from Ref. 173 with permissions.

Another intrinsic advantage of the nanorod metamaterial-based sensor is its porous texture, which not only tremendously increases the surface area for reaction, but also helps to achieve a large probing depth. Figure 12C displays the schematic of the flow cell, where the inset shows the field profile of the guided mode in cross-section. The nanorods are preliminarily immobilized with streptavidin molecules for functionalization. In the experiment, PBS buffer solution and biotin of different concentrations are pumped through the nanorod array (green wide arrows) and reflectivity is measured below the substrate with the ATR configuration. Figure 12D shows the real-time response of such a sensor to the reaction of biotin-streptavidin binding. Considering the small molecular weight of biotin, the fast and large resonant dip shift after the analyte injection is truly remarkable. This strongly indicates the power of the guided mode of nanorod metamaterials for biosensing. Noticeably, modern technique of nanofabrication is mature in processing nanorod/wire structures, which allows them to be integrated into complex systems for different applications and requirements.

**5.3 Plasmonic Metasurfaces**

The metasurface discussed in Section 3 could also open a new route for biosensing designs. Although SPPs are best known for their tremendous local field enhancement, the absolute reading of intensity variations has some intrinsic limitations in satisfying the high precision requirement of quantitative analysis. In practice, normally a spectral shift in resonance is more preferred as a metric parameter. Owing to the ability of causing anomalous reflection and refraction, the plasmonic metasurface can translate tiny refractive index changes at local positions into measurable offsets in the angle of the reflected or refracted beam.[77] In this way, one can produce ultrasensitive biosensors relying on the angle offset as schematically illustrated in Figure 13A.

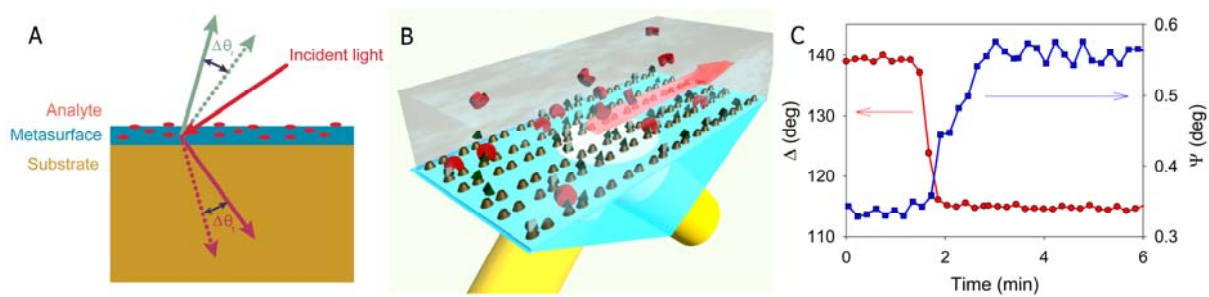

**Figure 13** Biosensing with plasmonic metasurface. (A) Analyte-adsorbed metasurface exhibits offsets in the angles of anomalous reflection and refraction, which is subject to the local index change. (B) Schematic of biosensing using a singular-phase nanodot array. (C) Evolution of ellipsometric parameters as biomolecules bind to the functionalized nanodots. The bonding of streptavidin to functionalized Au dots causes a quick change in the phase of reflected light at λ = 710 nm and an incidence angle of 53°. (A) is adopted from Ref. 77, and (B) and (C) are adopted from Ref. 177 with permissions.

Another measurable parameter that has been adopted in biochemical or biomedical detection is the phase of light. Specifically, it is the so-called "singular phase",[177] which refers to the extremely fast phase changes where the intensity of light becomes very close to zero. The corresponding detection can be realized with the help of plasmonic metasurfaces by measuring the reflected light beams. As it is demonstrated that the point of zero reflection exists in a variety of periodic nanostructures with only slightly different wavelengths, the design of the unit cells is flexible and not sensitive to their topologies. Figure 13B illustrates the schematic of biosensing based on a plasmonic double-nanodot array on a glass substrate. The array constant $a$ = 320 nm, average size of the dots $d$ = 135 nm, and dot separation of the pairs $s$ = 140 nm. To perform measurements, biotin is attached to the functionalized nanodots, which are represented by the dark green triangles and golden truncated cones respectively. The metasurface is exposed to streptavidin solutions, in which the target biomolecules are denoted by the red piechart-like disks. Light illuminates the array in the ATR geometry, and localized SPPs of the nanodots are coupled by the diffracted wave propagating along the surface, as shown by the red arrow. By tuning the wavelength and angle of incidence, the point of zero reflection can be achieved because of the "topological protection". Phase cannot be defined at the point of such complete darkness. However, at wavelengths close to this critical point, where the field intensity is still very small, sharp phase changes can be observed from the ellipsometric spectra. The experimental result for the setup in Figure 13B is reported in Figure 13C. The evolution of ellipsometric parameters, namely the amplitude $\Psi$ and phase $\Delta$ are plotted by the blue and red curves respectively. The attachment of streptavidin molecules to nanodots causes a steep drop of phase by ~25° within 2 min, which is much larger than the amplitude of the intensity change, promising an ultrasensitive real-time biomolecular recognition technique.

**5.4 Chiral Fields**
Chiral metamaterials are another important branch of metamaterials, which are normally constructed from chiral elements, such as 3D metallic helices[10] and 2D gammadion unit cells[201] and spirals.[202] The unique structure of chiral metamaterials leads to a series of interesting phenomena with high chirality, for example, generating superchiral electromagnetic fields.[203] This characteristic provides a direct connection to bio-applications. It is well known that many building blocks of life comprise chiral molecular units, including amino acids and sugars. The biomacromolecules formed from these units inherit and exhibit chirality at the molecular and supramolecular level. Superchiral fields can offer a strong interaction with chiral materials with one or two orders of magnitude larger than that of circularly polarized waves used in traditional chiroptical methods. Hence, progress in chiral metamaterials may pave an avenue to ultrasensitive characterization of chiral molecules, which is beyond the conventional detection techniques such as circular dichroism (CD), optical rotatory dispersion and Raman optical activity.

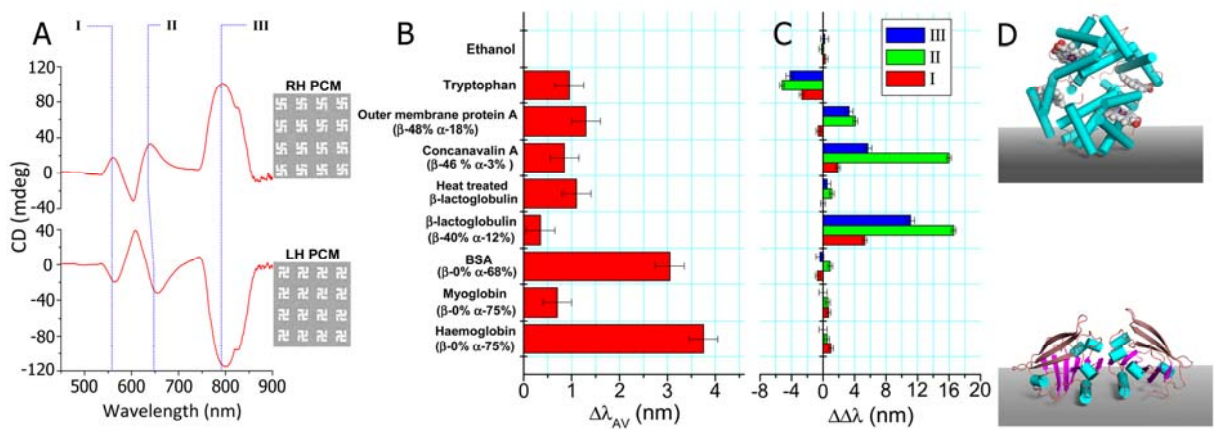

**Figure 14** Superchiral fields-based biosensing. (A) CD spectra of RH/LH PCMs. The curves for RH and LH gammadions are mirror images of each other due to their opposite chirality. (B) Average resonance shifts and (C) difference of resonance shifts between RH and LH PCMs on adsorption of chiral biomolecular layers. (D) Haemoglobin (top) and β-lactoglobulin (bottom) molecules adopt different geometries upon adsorption due to their different secondary structures (α-helix, cyan cylinders; β-sheet, ribbons). Figures are adopted from Ref. 174 with permissions.

An experimental study based on planar chiral metamaterials (PCMs) has been reported in Ref. 174, which employs left- and right-handed (LH/RH) gold gammadion arrays working at visible and near-infrared frequencies. The gammadion units can induce local field enhancement that is always desired for achieving larger interaction with biomolecules and better sensitivity to dielectric environment changes. Three resonant modes in the gold gammadion arrays are observed from the CD spectra in Figure 14A. In addition, the excited localized SPPs in LH/RH gammadions behave differently and are coupled between branches to create strong field chirality. This character is also reflected in the plot. As one can see from Figure 14A, the spectrum of the LH gammadions is a perfect mirror image of that of the RH ones, implying a reversal of chirality of the generated field. When a biomacromolecule layer is adsorbed to the surface of PCMs, the property of the resonance shifts induced by refractive index change is determined by the chirality of the molecules and that of the fields. For an achiral adsorption such as ethanol, almost no shift is observed according to the first line in Figures 14B and 14C, where the scales of the bars are defined as the average and difference of the chiral resonance shifts in RH and LH PCMs, respectively. In sharp contrast, the behaviors of these two parameters are found to be quite different for chiral molecular layers. Specifically, large dissymmetry is observed in the difference of shifts on adsorption of tryptophan (negative) and β-sheet proteins (positive), whereas the average shifts are larger for α-helical proteins. A special case is the heat-treated β-lactoglobulin. With the loss of β-structure at high temperature, the dissymmetry parameter exhibits a dramatic drop. The measurement results also provide indirect information about the geometry of the chiral molecules, which can be used to explain the different symmetric properties of the difference of resonance shifts. As can be seen in Figure 14D, biomacromolecules rich in α-helices (denoted by cyan cylinders) are more isotropically distributed at the interface with a relatively larger thickness, while the β-sheet proteins (represented by ribbons) result in anisotropic aggregation in lateral directions but a thinner structure along the surface normal. It was previously revealed that the largest contributions of chiroptical phenomena come from the interactions of the electric dipolar-magnetic dipolar and the electric dipolar-electric quadrupolar excitations of molecules. The dipolar contributions always take effect in different media, but the quadrupolar contributions average to zero in isotropic materials.[204] As a result, when superchiral field is generated around the PCMs, the steep local field gradients can dramatically enhance the dipole-quadrupole interactions over the dipole-dipole ones in β-sheet structures, leading to the large dissymmetries as in Figure 14C. Compared with schemes introduced in previous sections, a noticeable difference here is the measurement process only relies on the RH and LH PCMs but does not need any recognition protein, which could lead to a feasible technique in ultrasensitive bio-assaying.

# 6. Self-assembled Plasmonic Metamaterials

So far, various top-down fabrication methods, such as electron beam lithography, focused ion beam milling, direct laser writing and nanoimprint lithography, have been applied to fabricate plasmonic and metamaterial structures. Both planar and 3D geometries with feature size of sub-100 nm can be reliably realized with high yield. However, it still remains challenging to achieve the spatial resolution smaller than 10 nm in well-defined architectures based on top-down processing. Meanwhile, the non-parallel procedures in top-down fabrications generally lead to long fabrication time. In comparison, self-assembly approach promises low cost, high efficiency, precise structural control at truly nanometer scale, as well as sophisticated hierarchical structures with multifunction. Significant progress has been achieved in self-assembled plasmonic metamaterials in the past several years.

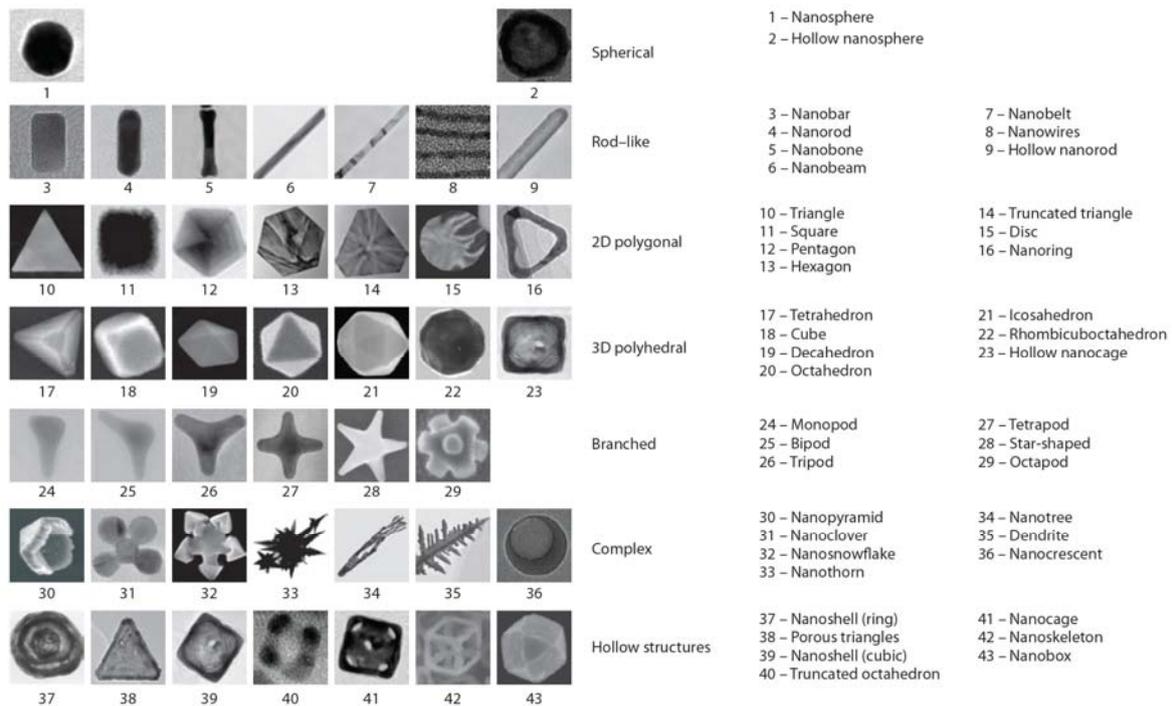

**Figure 15** Toolbox of metallic nanoparticles in different shapes, dimensionality and geometries, which serve as building blocks to self-assemble plasmonic metamaterials. The geometric order of the nanoparticles gradually increases from left to right in each row in terms of aspect ratios, number of sides, facets, or branches. Figure is adopted from Ref. 205 with permission.

The study of nanoscale self-assembly of metallic nanoparticles started more than two decades ago. Nowadays, high-quality metallic nanostructures with a variety of shapes (such as sphere, rod, cube, plate, pyramid, star, cage, core-shell, etc.) can be routinely achieved in laboratory through optimized wet chemical synthesis techniques. As shown in Figure 15, the geometry, facet or even inner structure of the nanoparticles could be rationally controlled, providing a complete, ingenious toolbox to construct complex metamaterials.[205] The interactions between nanoparticles, in addition to their size and shapes, play an important role in the resulting self-assembled geometry. These interactions include van der Waals forces, electrostatic forces, capillary forces, and molecular surface forces after decorating the surface of nanoparticles with various chemical functionalities.[206,207] The magnitude and direction (*i.e.*, attractive or repulsive) of these forces, as well as their working distances (*i.e.*, long range or short range) are distinctly different. The intricate balance of these interactions gives rise to the spontaneous organization of nanoparticles into a relative stable structure at a thermodynamic equilibrium state, where the Gibbs free energy of the entire system is minimum. It should be noted that the interparticle forces at the nanoscale are still not fully understood, although we have achieved considerable advances in the area of self-assembly.

In 2010, Capasso's group reported that self-assembled clusters of metal-dielectric core-shell structures exhibit pronounced electric, magnetic and Fano resonances, which can be readily tuned by adjusting the number and position of particles in the cluster.[208] The resonances of the clusters are attributed to the strong near-field coupling between closely spaced particles, which can be described by plasmon hybridization.[209,210] The building block of the clusters is gold-silica nano core-shell structures with tunable electric dipole resonance controlled by the core-shell aspect ratio. The spherical geometry supports orientation-independent coupling, ideal for constructing highly symmetric or isotropic metamaterials. Polymers with thiolated linkers in aqueous solutions are used to form monolayers on the surface of the core-shell nanoparticles. The polymer composition and chain length can be selected, leading to controllable gap size and optical coupling between nanoparticles. Finally, the polymer-coated nanoshells are assembled into clusters by slowly drying a droplet of the core-shell nanoparticles on a hydrophobic substrate (Figure 16A). Dark-field spectroscopy is applied to measure the intriguing optical properties of individual nanoshell clusters. For instance, a trimer, consisting of three core-shell nanoparticles, supports a magnetic dipole mode that originates from the induced in-plane current loop. In the experiment, a cross-polarizer is used with s-polarized incident light to filter out the scattered electric dipole radiation. This scheme is particularly useful for detecting the magnetic dipole mode that can be obscured by the broad electric dipole peak in scattering spectra. As shown in Figure 16B, the peak around 1400 nm corresponds to the magnetic dipole mode. The magnetic dipole mode exhibits a sharper spectral width and a red shift in comparison to the electric dipole mode, in good agreement with the simulation result. A 3D tetrahedral cluster can support isotropic magnetic resonances. Furthermore, Fano resonance is observed in a heptamer plasmonic cluster that is composed of seven identical elements. The overall dipole moment of the outer hexagon is similar in magnitude but opposite in sign with respect to the dipole moment of the central core-shell nanoparticle, leading to strong destructive interference of their radiating fields and the asymmetric line shape.

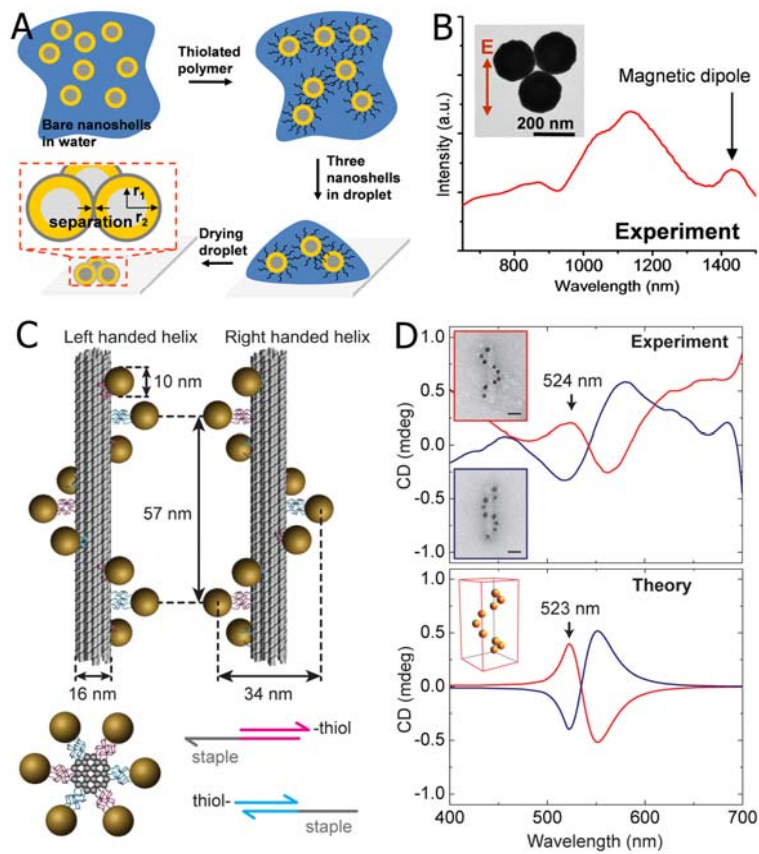

**Figure 16**  (A) Schematic illustration showing the procedures to assemble plasmonic clusters. (B) Scattering spectrum of a plasmonic trimer using cross-polarization method. The spectrum exhibits a clearly visible magnetic dipole resonance peak. The inset is the SEM image of the plasmonic trimer. (C) Schematic illustration showing DNA-guided self-assembly of plasmonic nanohelices. Left- and right-handed nanohelices

(diameter 34 nm, helical pitch 57 nm) are formed by nine gold nanoparticles each of diameter 10 nm that are attached to the surface of the DNA origami 24-helix bundles. (D) Experimental (top panel) and theoretical (bottom panel) circular dichroism spectra of the self-assembled gold nanohelices. The spectra of left-handed (red lines) and right-handed (blue lines) helices of nine gold nanoparticles show characteristic bisignate signatures. (A) and (B) are adopted from Ref. 208, and (C) and (D) are adopted from Ref. 217 with permissions, respectively.

DNA-guided assembly is an alternative, viable method for controllable arrangement of nanoparticles of different size and composition.[205] The unique and specific Watson-Crick base pairing of DNA, that is, adenine-thymine (A-T) and guanine-cytosine (G-C), offers systematically and precisely programmable hydrogen bonding. The length and the mechanical strength of DNA strands can be finely and widely tuned over a large range. All of these merits enable DNA as a superior interface to regulate the interactions between nanoparticles, and produce desired self-assembly hierarchy with judicious structural and sequence design. The recently developed DNA-origami technique even allows pattern designs in almost arbitrary manner.[211] The key is to use specific short single-stranded DNA as "staple" to direct the folding of a long single strand into the desired shape.

Utilizing DNA as a ligand through monofunctionalization and anisotropic functionalization, a vast assortment of plasmonic molecules and crystals has been realized. For example, the seminal work has shown that single-stranded DNA (ssDNA) can be attached to gold nanoparticles via a thiol linker, and subsequently nanoparticles are assembled by the hybridization of the DNA into a double helix.[212,213] A double-stranded DNA pyramidal scaffold is capable of controlling the position and arrangement of gold nanoparticles, resulting in discrete, tetrahedral nanostructures.[214] Such a tetrahedral plasmonic "molecule" exhibits strong chiral optical properties. Anisotropic functionalization provides another route to generate a diverse selection of discrete plasmonic molecules, in which the size and composition of individual nanoparticles could be varied.[215,216] DNA origami probably represents the ultimate solution for programmable and nanometer-precise design of plasmonic metamaterials. Using this technique, Kuzyk and co-workers fabricated chiral plasmonic nanostructures showing strong circular dichroism at visible wavelength.[217] Figure 16C is the schematic illustration of the design. Left- and right-handed nanohelices (helical pitch 57 nm) are formed by attaching nine gold nanoparticles (diameter 10 nm) to the surface of DNA origami 24-helix bundles (diameter 16 nm). Each of the nine attachment sites consists of three single-stranded extensions of staple oligonucleotides. Gold nanoparticles are coated with multiple thiol-modified DNA strands, which are complementary to these staple extensions. The mixture of gold nanoparticles and 24-helix bundles will hence result in the designed plasmonic helical structures. Due to their intrinsic geometric handness, left-handed and right-handed helices respond to left-circularly and right-circularly polarized light differently. The CD spectra, *i.e.*, the differential absorption of left-/right-circularly polarized light, are plotted in Figure 16D. The characteristic bisignate peak-dip line shapes, which are vertically mirrored for the two helical structures, agree well with theoretical calculations in terms of the spectral wavelength and signal strength. The CD signal can be further enhanced when nanoparticles are larger or arranged in a smaller helical pitch.

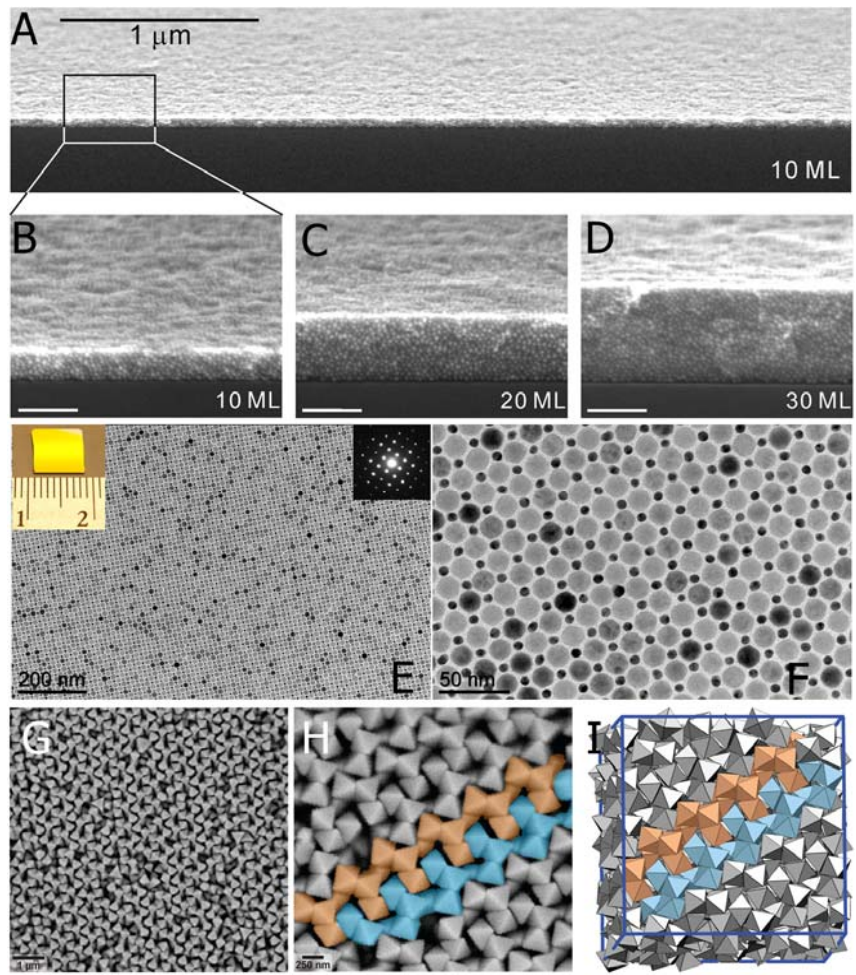

**Figure 17** (A-D) SEM micrographs of layer-by-layer assembled 3D supercrystals made of Au nanoparticles (diameter 6 nm). (A) SEM micrographs of an Au nanoparticle supercrystal with 10 monolayers over a large

area. (B-D) High-resolution cross-sectional SEM micrographs of Au nanoparticle supercrystals of 10, 20 and 30 monolayers, respectively. (E) TEM image of an AB-type binary superlattice monolayer self-assembled from 16.5 nm $Fe_3O_4$ and 6.4 nm Au nanoparticles. The upper left inset shows a photograph of a $SiO_2$/Si wafer coated with such a superlattice, and the upper right inset shows the small-angle electron diffraction pattern. (F) High-magnification TEM image of the monolayer in (E). (G-I) Self-assembled Ag octahedra with a previously unknown lattice in the presence of excess PVP. (G) SEM micrograph showing the new octahedron supercrystal. The lattice consists of tetramer motifs (H) and forms spontaneously in Monte Carlo simulations of octahedra with depletion attractions (I). (A-D) are adopted from Ref. 222, (E) and (F) are adopted from Ref. 226, and (G-I) are adopted from Ref. 228 with permissions.

In terms of large-area self-assembled plasmonic metamaterials, different approaches have been proposed and demonstrated.[218-221] For example, Gwo's group reported a simple method for synthesizing 3D plasmonic crystals over an area >1 $cm^2$ consisting of gold and silver nanoparticles (Figures 17A-17D).[222] In their experiments, Janus nanoparticles, which have solvent-phobic top surface and solvent-philic bottom surface, form close-packed monolayers based on layer-by-layer (LbL) assembly from suspensions of thiolate-passivated gold or silver colloids. Such plasmonic crystals exhibit strong transverse (intra-layer) and longitudinal (inter-layer) near-filed coupling. Furthermore, large-area superlattices with two or three types of nanostructures promise a new route for designing metamaterials with rationally tunable properties by adjusting the chemical composition, size, and stoichiometry of the constituents.[223,224] Dong *et al.* recently demonstrated the self-assembly and transfer of macroscopic, long-range-ordered supperlattices made of binary, or trinary nanocrystals using the liquid air interfacial assembly approach (Figures 17E and 17F).[225,226] New synthetic methods result in monodisperse Ag nanocrystals with well defined polyhedral shapes, including cubes, truncated cubes, cuboctahedra, octahedra and truncated octahedra.[227] The particle shape provides an additional handle to twist the structure of assembled patterns, when the polyhedra are self-assembled into long-range ordered superstructures. In one experiment, monodisperse Ag polyhedra, coated with polyvinylpyrrolidone (PVP), assemble into large, dense supercrystals by gravitational sedimentation.[228] Unlike cubes and truncated octahedra, octahedra in their densest lattice packing show only incomplete face-to-face contact, and the equivalent faces of neighbouring octahedra either do not lie in a common plane. Unadsorbed PVP polymer in solution will induce attractions between particles, leading to new structures that are not governed solely by packing efficiency of hard shapes. As shown in Figures 17G and 17H, a lattice of octahedra with intriguing helical motifs is generated, if extra PVP is added to the solution immediately before sedimentation. Higher amounts of PVP result in less Minkowski lattice and more helical lattice. This characteristic is confirmed by Monte Carlo simulation (Figure 17I). The large degree of face-to-face contact in the latter reflects significant attractive forces.

## 7.     Conclusion and Outlook

We have seen rapid growth in the field of plasmonic metamaterials, and it will continuously advance. The integration of plasmonics and metamaterials results in several benefits, rendering plasmonic metamaterials a promising platform to explore complex optical effects and new applications. First, the large local field enhancement leads to stronger light-matter interaction, through which many weak processes can be enhanced and effectively utilized. Second, the tight field confinement and ultrafast optical process enable the construction of miniaturized devices, working either as the bridge of photonic and electronic components, or even all-optical photonic circuits. Third, the spectra of plasmonic metamaterials are highly tunable by altering the responses of individual constituents and their coupling. This is highly desirable in almost all the practical situations. Owing to these attractive features, plasmonic metamaterials are expected to extend to a much broader horizon beyond what we have discussed in this review article. We think that the following areas are still largely unexplored, and worth to devote special efforts.

(1)     The first one is nonlinear optics in plasmonic metamaterials. The large local field can provide strongly enhanced nonlinear processes, if we embed nonlinear materials into plasmonic nanostructures. In addition, the metal itself has high intrinsic nonlinearity, which can be further amplified by texturing the metal surface. An enhancement of four orders of magnitude in conversion efficiency for optical second-harmonic generation has been experimentally demonstrated in a periodically nanostructured metal structure.[229] Very recently, up to the 43$^{rd}$ harmonic generation of ultrashort extreme-ultraviolet pulses have been achieved by means of the plasmonic field enhancement.[230,231] It is shown that the strong magnetic resonance, whose contribution is normally ignored, could substantially modify the nonlinear optical process.[232,233] In terms of negative refraction, an alternative approach based on phase conjugation from a thin nonlinear film has been proposed.[234,235] Such a scheme was experimentally demonstrated in the microwave region,[236] and at optical frequencies using four-wave mixing in nanostructured metal films[237] and graphite thin films,[238] respectively. Other intriguing nonlinear phenomena, such as unusual phase matching[239,240] and solitary wave propagation[241,242] are worth for experimental exploration. All of these exciting phenomena and developments promise tunable metamaterials, ultrafast switching and optical modulation.[243-246]

(2)  There is an explosively growing interest in expanding the properties of plasmonics and metamaterials towards the quantum regime, and studying how plasmonics and metamaterials can mediate light-matter interaction with new functionalities at the nanoscale.[247] For example, researchers have demonstrated the survival of photon entanglement in plasmonic systems,[248] the blue shift and line-width broadening of the plasmon resonance due to the quantum size effect,[249,250] and electron tunneling between two nanoparticles with a separation distance less than 1 nm.[251,252] New theories based on the non-locality[253,254] and the quantum-corrected model[255] have also been developed to study the plasmonic field enhancement at such extremely small scales. In addition, the large photonic density of states in hyperbolic metamaterials is able to increase the interaction with the quantum emitter while simultaneously channeling the light into a subdiffraction single-photon resonance cone.[256] The research outcome in the new frontier of quantum plasmonics and quantum metamaterials will considerably advance our understanding of the fundamental physics associated with SPPs and metamaterials, and facilitate the development of novel quantum photonic and electronic components,[257-261] including single-photon sources, nanoscale lasers, transistors, and detectors.

(3)  Dynamically reconfigurable plasmonic metamaterials is also a valuable subject to explore. Although the modulation of subwavelength features at optical regime is challenging, significant advancement has been achieved to overcome this obstacle. The usage of graphene, phase-change media,[262,263] liquid metals[264] or liquid crystals[265,266] provides possible solutions from the material aspect, and electrical modulation of structural patterns[267] and MEMS[268,269] solves the problem from the structure aspect. In particular, plasmonic metamaterials operating in the fluidic environment may realize the ultimate reconfigurability and multiple functionalities. Very recently, laser-induced surface bubbles are used to manipulate the propagation of SPPs. By precisely controlling the size, location and shape of the surface bubbles, as well as the phase front of the incident SPPs, divergence, collimation, and focusing of SPPs are all obtained by such plasmofluidic lenses.[270] Moreover, molecular-controlled nanoparticle assembly in solutions has been successfully demonstrated, providing a powerful building block of plasmonic metamaterials at deep sub-wavelength scale.

In conclusion, we have witnessed tremendous progress of metamaterials and plasmonics over the past decade. The alliance of metamaterials and plasmonics will inevitably offer novel opportunities to advance both fields and stimulate new cross-disciplinary approaches to address grand challenges in information processing, clean energy, and human health. Many important discoveries and applications in the emerging field of plasmonic metamaterials are awaiting us to exploit.